\documentclass[twocolumn,floatfix]{revtex4}
\usepackage[usenames]{color}
\usepackage{amsmath} 
\usepackage{amssymb} 
\usepackage{graphicx} 
\usepackage{epsfig}  
\usepackage{ulem}

\begin{document} 
\title{Novel chiral quantum spin liquids in Kitaev magnets} 
\author{Arnaud Ralko} 
\affiliation{Institut N\'eel, UPR2940, Universit\'e Grenoble Alpes et CNRS, Grenoble 38042, France} 
\author{Jaime Merino} 
\affiliation{Departamento de F\'isica Te\'orica de la Materia Condensada,
Condensed Matter Physics Center (IFIMAC) and Instituto Nicol\'as Cabrera,
Universidad Aut\'onoma de Madrid, Madrid 28049, Spain} 
\date{\today} 

\begin{abstract}
Quantum magnets with pure Kitaev spin exchange interactions can host a gapped
quantum spin liquid with a single Majorana edge mode propagating in the
counter-clockwise direction when a small positive magnetic field is applied.
Here, we show how under a sufficiently strong positive magnetic field a
topological transition into a gapped quantum spin liquid with two Majorana edge
modes propagating in the clockwise direction occurs. The Dzyaloshinskii-Moriya
interaction is found to turn the non-chiral Kitaev's gapless quantum spin
liquid into a chiral one with equal Berry phases at the two Dirac points.
Thermal Hall conductance experiments can provide evidence of the novel
topologically gapped quantum spin liquid states predicted.
\end{abstract}
\maketitle 
 
A quantum spin liquid (QSL) is an exotic state of matter in which localized
spins do not order even at zero temperature in contrast to magnetic ordering
observed in conventional insulating magnets. QSL's are highly entangled states
which cannot be characterized by a Landau local order parameter. Exhibiting
topological order, emergent gauge fields and fractional excitations
\cite{balents,savary}, they are at the heart of an intense research activity.
The first concrete example of a two-dimensional QSL has been the resonating
valence bond (RVB) state envisioned by Anderson \cite{anderson1973} to describe
the ground state of triangular antiferromagnets and as the parent insulating
phase of high-T$_c$ superconductors. Due to the spin correlations, a spin-flip
in an RVB state fractionalizes into two neutral spin-1/2 particles (spinons)
which can propagate freely around the lattice. In spite of intense experimental
efforts, there is no unambiguous observation of fractionalization -- such as
the expected spinon continuum in the spin excitation spectra -- in real
materials.

The interplay between strong Coulomb interaction and spin-orbit coupling
\cite{jackeli} in the honeycomb magnets such as A$_2$IrO$_3$~\cite{takagi} (with
A=Li, Na), H$_3$LiIr$_2$O$_6$ \cite{kitagawa}, $\alpha$-RuCl$_3$, and
organometallic frameworks \cite{oshikawa2017} can lead to special compass
interactions which frustrate the magnetic order of the $S=1/2$ pseudospins. The
exact QSL ground sate of the Kitaev \cite{kitaev} model has opened the
possibility of finding fractionalized excitations in spin-orbit coupled Mott
insulators on honeycomb lattices. From the decomposition of the spin operators
onto four non-interacting Majorana fermions, Kitaev showed that the elementary
spin excitations of the Kitaev QSL (KQSL)  are fractionalized into itinerant
Majorana fermions with Dirac dispersion, and localized ones giving Z$_2$ gauge
fluxes. Recent observations on $\alpha$-RuCl$_3$ \cite{motome,ruegg} and
H$_3$LiIr$_2$O$_6$ \cite{kitagawa} have been interpreted in terms of the
existence of such two types of excitations. 

However, a realistic description of honeycomb materials requires including
additional spin interactions not included in the Kitaev model as well as
considering large magnetic fields beyond the perturbative regime discussed so
far.  Since there is no exact solution in these physically relevant situations
new theoretical approaches are required to properly describe the system. For
instance, exact numerical and slave fermion approaches
\cite{yfjiang,hcjiang,lzou} of the Kitaev model have found a transition to a
gapless $U(1)$ spin liquid phase under sufficiently strong applied magnetic
fields \cite{trebst,kimchi}. On the other hand, the effect of Heisenberg and
symmetric spin exchange terms needs to be considered~\cite{schaffer,knolle}
in order to accurately describe the magnetically ordered phases \cite{gholke}
observed in Na$_2$IrO$_3$ and $\alpha$-RuCl$_3$. Finally, the
next-nearest-neighbor Dzyaloshinskii-Moriya (DM) has been invoked as being
relevant for the description of real materials \cite{valenti} but its effect on
the Kitaev model remains little explored so far \cite{lunkin}.

Here, we report on two novel topological QSLs arising when either a strong
magnetic field or a DM interaction are considered in the pure Kitaev model.  We
have discovered that the gapped QSL state with Chern number $\nu=\pm 1$
(depending on the direction of the field) predicted at low tilted magnetic
fields \cite{kitaev}, undergoes a novel topological transition to a different,
topologically gapped QSL with $\nu=\pm 2$.  Such topological transition occurs
in a regime in which the Dirac cones disappear due to strong hybridisation
between itinerant and localized Majorana fermions.  We also predict the
presence of a novel gapless chiral QSL induced by the DM interaction and that
is characterized by equal Berry phases at the two Dirac cones
($\phi_K=\phi_{K'}= \pm \pi$) in contrast to the opposite Berry phases found in
the pure Kitaev model ($\phi_K=-\phi_{K'}= \pm \pi$). The novel topological
gapped QSL states found here could be tested through thermal Hall experiments.

%
Since we are interested in the description of competing topological phases
starting from the KQSL, the $H_K = 2 \sum_{\langle ij \rangle, \gamma} K^\gamma
S^\gamma_i S^\gamma_j$ -- where the three nearest-neighbor bonds $\langle ij
\rangle $ of the honeycomb lattice are denoted by $\gamma=x,y,z$ -- is the most
relevant starting point \cite{RemarkOnHeisenbergTerm}. We
emphasize the factor of 2 in the definition of the model.
The next-nearest-neighbor DM interaction: $H_{DM} =
\sum_{\langle \langle ij \rangle \rangle} {\bf D}_{ij}\cdot {\bf S}_i \times
{\bf S}_j$ is also important for describing the magnetic orders observed in
Iridates \cite{valenti,perkins}, and it is known that, combined with the
magnetic field,  can open a non-trivial topological gap. For this purpose, we
also consider the term $H_B = -\sum_{i} {\bf B} \cdot {\bf S}_i$. Hence, the
final hamiltonian reads: 
\begin{eqnarray}
H    &=& H_K + H_{DM} + H_B, 
\label{eq:model}
\end{eqnarray}
and we will consider an isotropic Kitaev interaction {\it i.~e.} $K^\gamma=K$
in the rest of the paper, as well as a the magnetic field ${\bf B}$ and the DM interaction ${\bf D}_{ij}={\bf D}$ parametrized in function of tilt parameters $t$ and $d$ as $ {\bf B} = B ( t ,
t , 1 ) / \sqrt{1 + 2 t^ 2} $ and ${\bf D}=D(d,d,1)/ \sqrt{1 + 2 d^ 2}$
respectively, both ranging from the pure $z$ direction ($t$ or $d=0$) to the case perpendicular to
the honeycomb plane ($t$ or $d=1$).

\begin{figure}[ht]
\includegraphics[width=0.4\textwidth,clip]{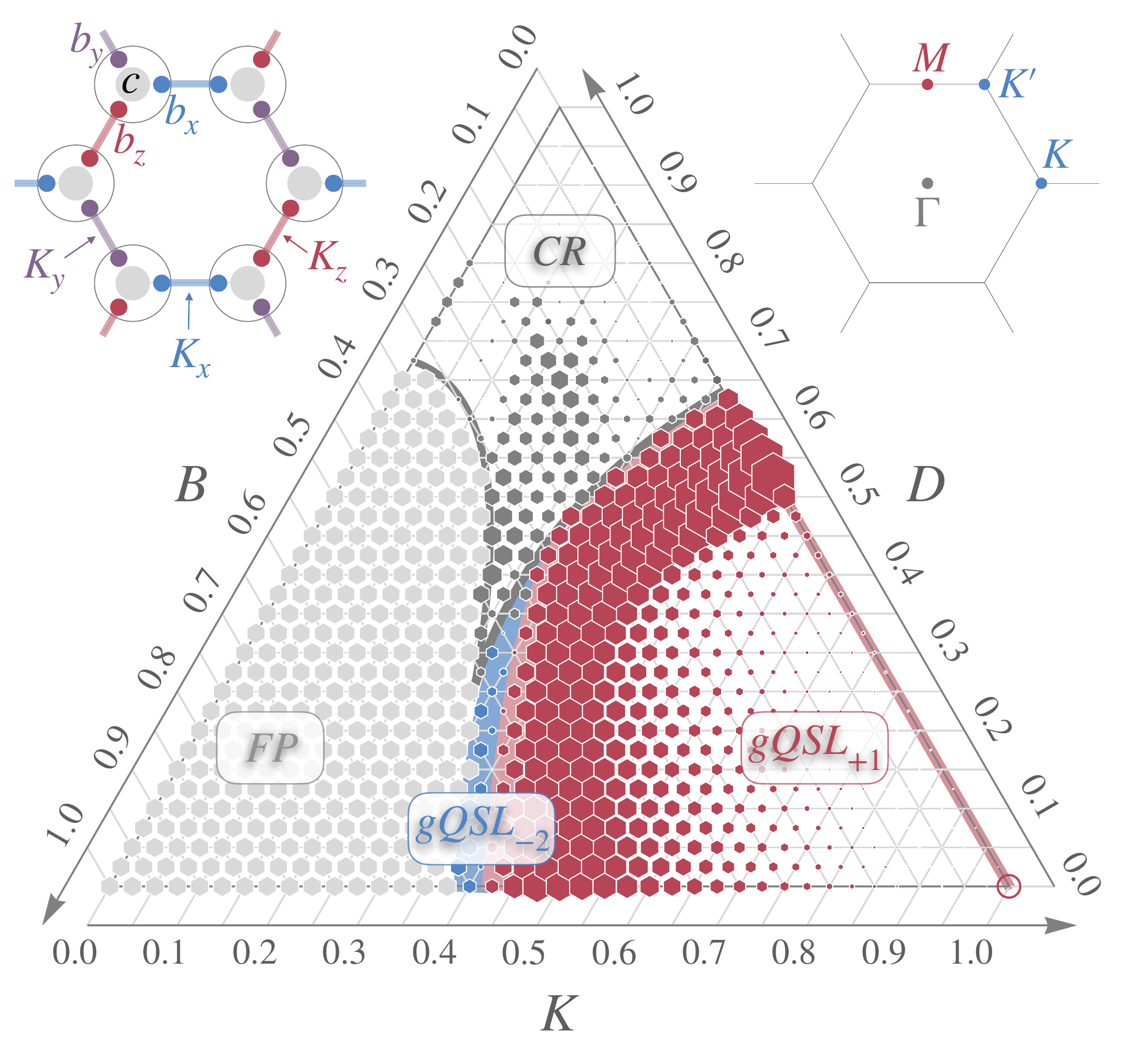}
\caption{
		Phase diagram of Kitaev model with DM interaction under magnetic field.
		The full phase diagram of the Kitaev model in the presence of a
		magnetic field pointing in the $[1,1,1]$ direction ($t=1$) is shown.
		Empty regions indicate gapless phases and the size of the hexagonal
		symbols indicate the size of the gap. Gray areas correspond to the
		gapped fully polarized (FP) phase with Chern number $\nu=0$, the red to
		a gapped QSL with $\nu=+1$ denoted as gQSL$_{+1}$. A gapped QSL with an
		unconventional Chern number of $\nu=-2$, termed gQSL$_{-2}$ (dark blue)
		occurs between the FP and the gQSL$_{+1}$. Away the
		KQSL at $K=1$ (open circle), an ungapped QSL, uQSL
		(thick red line) with equal Berry phases at the two
		Dirac points occurs for $B=0$ and $K=1$ at a non-zero DM, $D<0.5$,
		which becomes the gQSL$_{+1}$ around $D \sim 0.5-0.65$, to finally
		become gapless and non- topological at a larger $D$ (dark gray). In
		this region, CR refers to classical regimes, beyond the accessibility
		of the present theory. The left inset shows the Majorana decomposition
		of the model considered while the right inset shows the first Brillouin
		zone and symmetry points of the honeycomb lattice.} \label{fig:fig0}
\end{figure}

The full model is treated using Kitaev's Majorana decomposition of the spin
$S^\alpha = \frac{1}{4}( i b^{\alpha} c - i \frac{1}{2} \epsilon_{\alpha,
\beta, \delta} b^{\beta} b^{\delta} )$ \cite{kitaev} where greek letters span
the space dimensions $(x,y,z)$ and $\epsilon_{\alpha, \beta, \delta}$ is the
Levi-Civita symbol (Fig.~\ref{fig:fig0}). In this notation, $b$ operators
correspond to flux band variables while $c$ describes the itinerant Majorana
fermions. These  are in the presence of the three other localized Majoranas
which act as gauge fluxes (see Fig. 3(a) for the dispersion relation of the
pure Kitaev model). 

Going away from the specific Kitaev point requires the proper consideration of the
constraint on the number of fermions -- four Majoranas per spin -- that can be
only achieved in average by introducing Lagrange multipliers $\{ \lambda_i \}$:  
\begin{eqnarray*}
	H_L &=& \frac{i}{4}\sum_{i, \alpha} \lambda_\alpha \left[ b^\alpha_i c_i +
\frac{1}{2}\epsilon_{\alpha \beta \delta } b^\beta_i b^\delta_i \right]
\label{eq:lagr}
\end{eqnarray*}
where 
summation over repeated indices is assumed. The explicit implementation of the 
fermion constraint is crucial for a proper description of model (\ref{eq:model}) 
containing terms other than the pure Kitaev contribution and we have provided all details in \cite{suppl}.

The complete phase diagram of model (\ref{eq:model}) can be nicely represented
in a ternary plot as displayed in Fig. \ref{fig:fig0} for ${\bf B}$ in the
$[1,1,1]$ direction ($t=1$) and ${\bf D}$ along the $z$-direction ($d=0$) for
simplicity, realizing that the case $d=1$ is qualitatively similar.
On this graph only, the full parameter range of the model onto the plane fulfills
the constraint $K+D+B=1$, providing $K>0$, $D>0$ and $B>0$, the area of the
hexagons is proportional to the gap at this point, and the color refers to
different states of matter. In this ($K,D,B$) constrained space, the vertex defined by
(0,0,1) corresponds to the topologically trivial ($\nu=0$) fully polarized (FP)
state, (1,0,0), to the gapless Kitaev QSL (KQSL) and (0,1,0) to a gapless
classical state whose magnetic properties remain yet to be determined. 
Three different topological phases characterized by their Chern numbers can be
distinguished in the phase diagram. A gapped topological QSL with $\nu = +1$
denoted by gQSL$_{+1}$ is topologically equivalent to the QSL found by Kitaev
at weak magnetic fields.  The novel gapped QSL with unconventional Chern number
of $\nu = -2$ is denoted by gQSL$_{-2}$. The gray areas correspond to the FP
state with $\nu =0$. 
The mechanisms driving these topological states are explained below,
but we emphasize here the presence of the two novel gapped topological QSL
with large Chern numbers, $\nu=\pm 2$. These are the result of the strong competition
between the three terms entering the hamiltonian and occur in intermediate
regions between the FP and the gQSL$_{+1}$ phases. These phases are {\it
chiral} QSLs since time reversal symmetry is broken either explicitly by the applied 
magnetic field or spontaneously for $B=0$ and $D \neq 0$.
Interestingly, at zero field, $B=0$, a small but finite DM of $D \lesssim 0.5$ (with $K=1$) 
leads to equal Berry phases, $\phi_k$, around the Dirac nodes (see Fig. \ref{fig:fig0} for the
first Brillouin zone of the lattice) $\phi_K=\phi_{K'}= \pm \pi$ in
contrast to the opposite Berry phases found in the pure Kitaev model, $\phi_K=-\phi_{K'}= \pm \pi$. 
Hence, we have unraveled a new ungapped QSL with non-zero chirality which we 
denote by uQSL.

Applying a tilted
magnetic field is found to open a gap in the Majorana fermion spectrum,
consistent with perturbation theory \cite{kitaev}. The gap is opened
symmetrically with respect to the zero energy and the resulting gapped QSL is
topological with a nonzero Chern number $\nu=\pm 1$, the sign depending on the
direction of the magnetic field.
Our gap arises naturally from imposing the constraints on the Majorana fermions in contrast to a previous Majorana mean field analysis 
which add by hand a three-spin term to be able to open a gap \cite{nasu}. We 
associate the gap opening to the non-zero Lagrange multipliers (\ref{eq:lagr}) which lead to 
local hybridization between matter and flux Majorana fermions for $B \neq 0$. We note that the one-particle constraints are not 
automatically verified when considering models beyond the pure Kitaev model even at the mean-field level
so it is necessary to impose them explicitly \cite{suppl}. Hence, our MMFT is capable of describing correctly the exact KQSL at $B=0$ as well as the gapped QSL under low 
magnetic fields without making any extra assumptions. We finally note that  closely related Abrikosov fermion mean-field theories \cite{gholke} imposing
the one single-particle constraint through a single Lagrange multiplier \cite{liu} (our $\lambda_z$) instead of three \cite{choi} -- as we have done --
can lead to different results to our MMFT~\cite{suppl}.

\begin{figure}[ht]
\includegraphics[width=0.45\textwidth,clip]{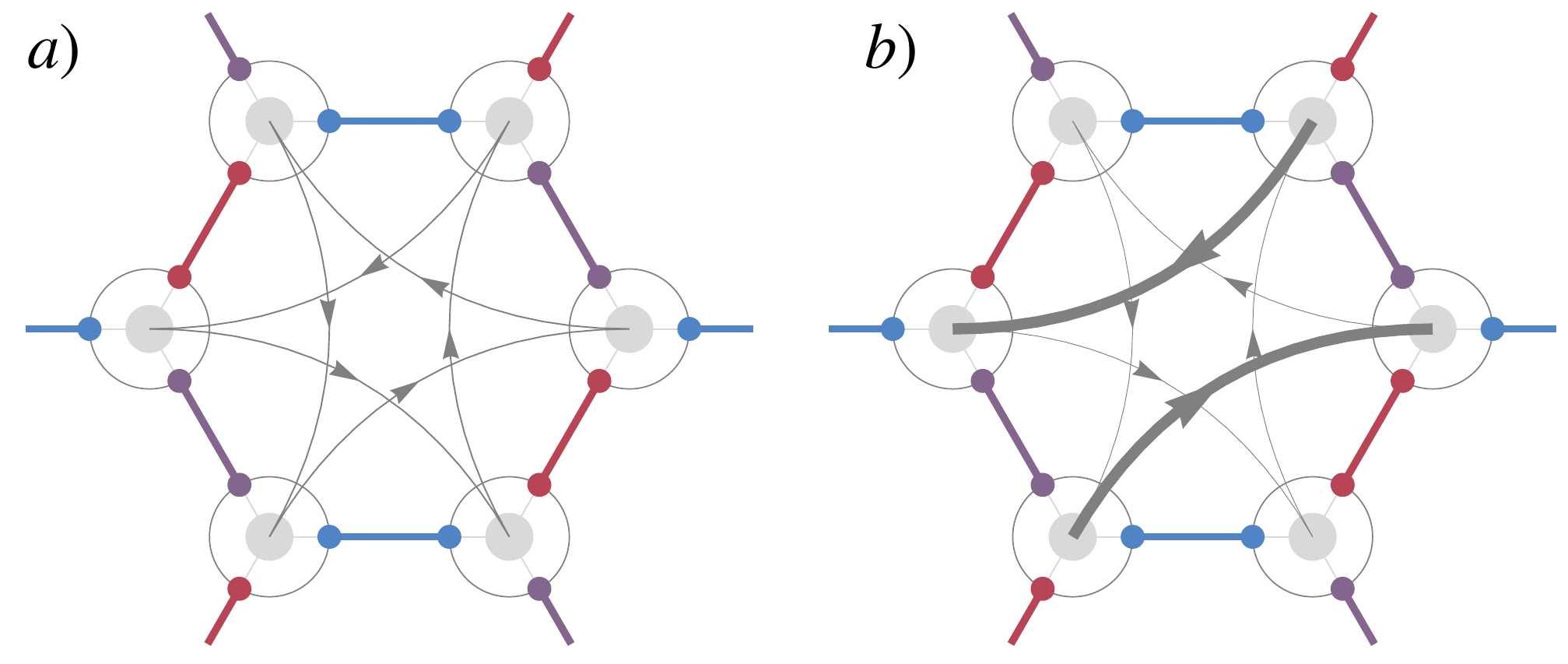}
\caption{Chiral amplitudes in the Kitaev model from the
Majorana mean-field theory. In (a) we show the  n.n.n. chiral
Majorana amplitudes, $\langle c_i c_j \rangle $, induced in
the ground state MMFT wavefunction either by a magnetic field perpendicular to 
the plane, $t=1$, or a DM vector oriented perpendicular to the plane ($d=1$).
These are responsible for the {\it chiral} QSL with Chern number $\nu=+1$
(gQSL$_{+1}$) discussed in the text. (b) Anisotropic amplitudes
(the thickness corresponds to the different strength) arising from finite DM $d\ne 1$
(with no applied magnetic field) responsible for the opening of the gap for
$0.5 <D <0.65$ (with $K=1$ and $d=0$) in the gQSL$_{+1}$.  The snapshots
show one of the two possible chiralities of the ground state ($\nu=+1$). Note
that (a) and (b) amplitudes are smoothly connected as shown in the phase
diagram of Fig.~\ref{fig:fig0}.
} 
\label{fig:fig1}
\end{figure}

Concomitantly with the gap opening, the magnetic field leads to non-zero chiral currents of Majorana 
fermions between the n.n.n sites as shown in Fig.~\ref{fig:fig1} (a).  Only when the three components of the magnetic field are
non-zero, all $\lambda_\alpha$ are simultaneously non-zero as well. This leads
to hybridization among the four Majoranas at each site which ultimately leads
to the gap opening.  The size of the gap depends on $t$, the actual orientation
of the  magnetic field. When the field is parallel to one of the natural
axis (say $t=0$ for ${\bf B}=B u_z$), the gap is zero and we have a 
gapless QSL. At a critical $B$ we find that the $\pm \pi$ Berry phases at the Dirac points
can switch their signs as found earlier \cite{nasu}.

We now discuss the effect of DM on the pure Kitaev model ($B=0$). As stated above, as the DM is increased the Berry phases around the Dirac points 
become equal: $\phi_K=\phi_{K'}=\pm \pi$, indicating a change in the nature of the KQSL which characterizes the uQSL.
As the $D$ parameter is further increased above a critical
value, $D \gtrsim 0.5$ (with $B=0$ and $K=1$), the system opens up a gap leading
to a topologically gapped {\it chiral }QSL with $\nu = \pm 1$ for either $d=0$ or $1$, gQSL$_{\pm1}$.
The origin of the non-zero Chern number may be associated with the
occurrence of {\it anisotropic} chiral amplitudes between n.n.n. sites as shown in
Fig.~\ref{fig:fig1}(b) for $d <1$. This lattice nematicity induced by anisotropy of  ${\bf D}$ is completely restored for $d=1$ 
at which the n.n.n. chiral amplitudes are the ones displayed in (a). In any case, the DM interaction induces both an {\it ungapped} and a {\it gapped}
phase. This uQSL is a novel QSL which breaks TRS spontaneously in contrast with 
the {\it gapped} chiral QSL reported on the decorated honeycomb lattice \cite{yao} 
due to its gapless character.
\begin{figure}[ht]
\includegraphics[width=0.45\textwidth,clip]{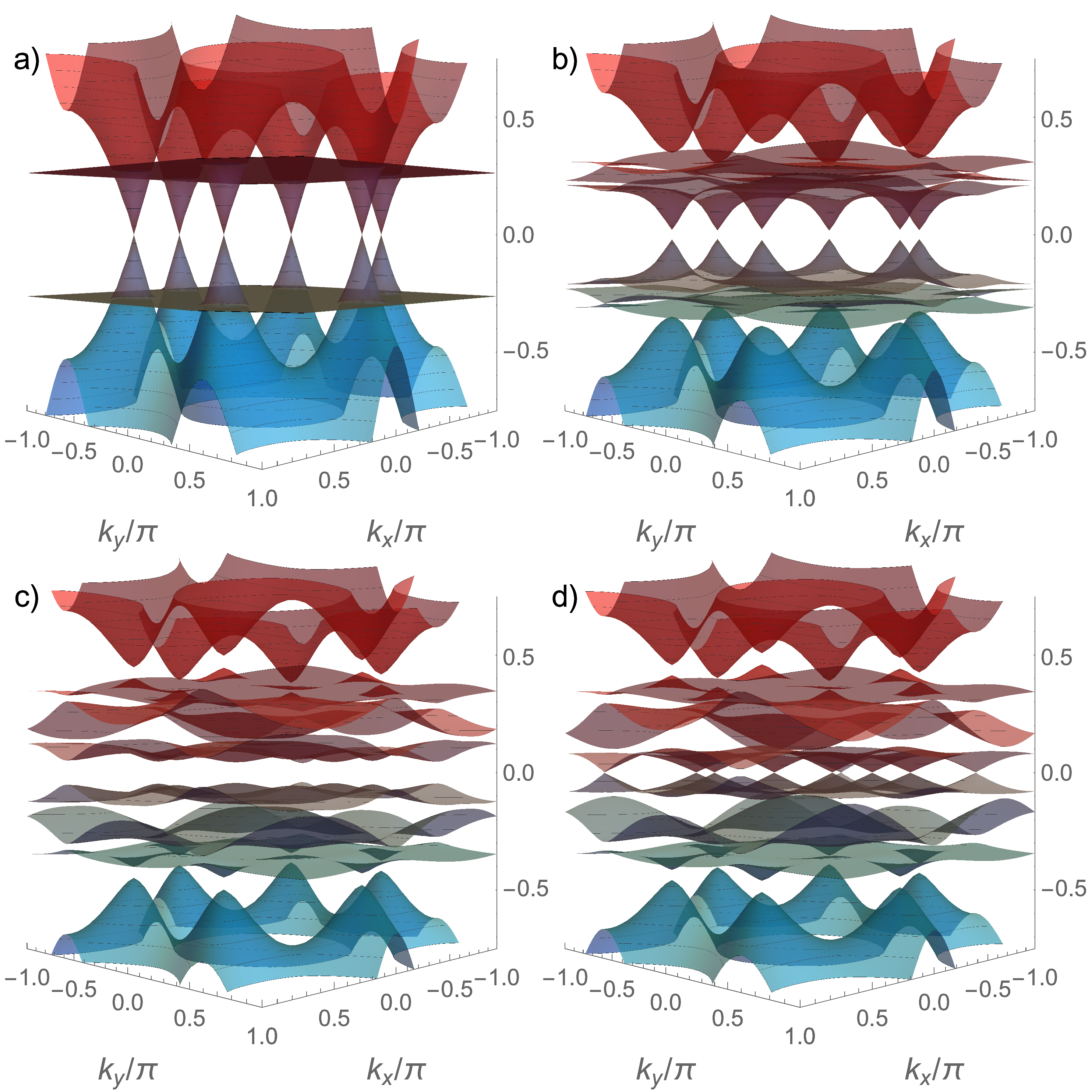} 
\caption{Majorana dispersions of the
Kitaev model under a tilted external magnetic field and zero DM interaction. The evolution of the
Majorana dispersions under a tilted $t=1$ magnetic field comparing the case (a)
with no applied magnetic field, $B=0$, consisting of gapless Majorana
dispersions and flat flux bands (three-fold degenerate) describing the Z$_2$
fluxes, (b) with $B \simeq 0.6$, where a gap has already opened and the flux
Majorana bands remain flat. This gapped QSL with Chern number $\nu=+1$,
gQSL$_{+1}$, persists up to $B_c=1.43$ is adiabatically connected with the KQSL, 
(c) with $B = 1.2$, the Dirac cones have been
washed out and the gapped fluxes are strongly distorted becoming dispersive, (d) with $B=1.43$, a gapped QSL with $\nu=-2$,
gQSL$_{-2}$, arises between $B_c$ and $B \simeq 1.64$ indicating a different
topological state from the KQSL. $K \equiv 1 $ in this plot.} 
\label{fig:fig2}
\end{figure}
When a $[001]$ magnetic field ($t=0$) is applied in the
presence of a nonzero DM, a gapless QSL with the two Dirac cones shifted in
opposite directions by the same amount occurs. The sublattice symmetry 
respected by the DM interaction protects Dirac cones from opening a gap. By tilting the magnetic field to
$t=1$,  a gap opens up as found in the case of zero DM leading to a gQSL$_{\pm
1}$. But unlike opening symmetrically around zero, the gap centers are equally shifted in
opposite directions at the two cones.
\\
In Fig.~\ref{fig:fig2} we show the evolution of the Majorana dispersions under
a $[111]$ magnetic field ($t=1$). With no magnetic field applied the MMFT
dispersions consist on gapless Majorana matter bands touching at the Dirac
points and flat bands describing localized Z$_2$ fluxes. The flat bands are
three-fold degenerate in this case. As the magnetic field increases up to about
$B \sim 0.6$, (with $K=1$) a gap opens up at the Dirac points while the flux
bands remain almost  flat. This gQSL$_{+1}$ phase -- since it
is gapped and $\nu=+1$ -- is adiabatically connected to the gapped QSL
found by Kitaev \cite{kitaev}.
As the magnetic field is increased up to
$B_c\simeq1.43$ the flux bands are gradually distorted
becoming dispersive and strongly hybridized with the Majorana matter bands.
Note also how the bands are becoming closer at the three $M$ points for $B \sim B_c$.
As the field is increased beyond $B_c$, a gap opens up at this newly formed
band touching $M$ points with Berry flux magnitudes larger than $\pi$ in contrast
to typical Dirac cones. This can be attributed to the fact that the matter and flux Majoranas are now forming 
unseparable composite objects due to the strong hybridization in this magnetic field regime~\cite{suppl}.
Hence, we find a gapped topological QSL with Chern number $\nu =-2$ emerging in
the range $B \in [1.43,1.64]$. 
It is interesting to note that a CSL with $\nu=2$ has been found in \cite{liu} but with an extra symmetric spin term
in the hamiltonian.
At larger magnetic fields ($B>1.64$) there is a
transition to a gapped and fully polarized insulator  with a trivial topology
($\nu=0$) -- the full polarization is an artefact of the method. The Majorana dispersions are also strongly modified by $D$ even for
$B=0$.  Recent numerical studies \cite{yfjiang}-\cite{trivedi} suggest the
existence of a {\it gapless} intermediate phase in a somewhat similar parameter
range. In spite of the different features found (gapless vs. gapped), our gap
closing around $B_c$ can be associated with the large enhancement of low energy
excitations developing near the PL phase. Interestingly, we have supported by
ED the fact that $D$ and $B$ combined  possibly lead to a gap
opening.  All these points are detailed in \cite{suppl}.

\begin{figure}[ht]
\includegraphics[width=0.35\textwidth,clip]{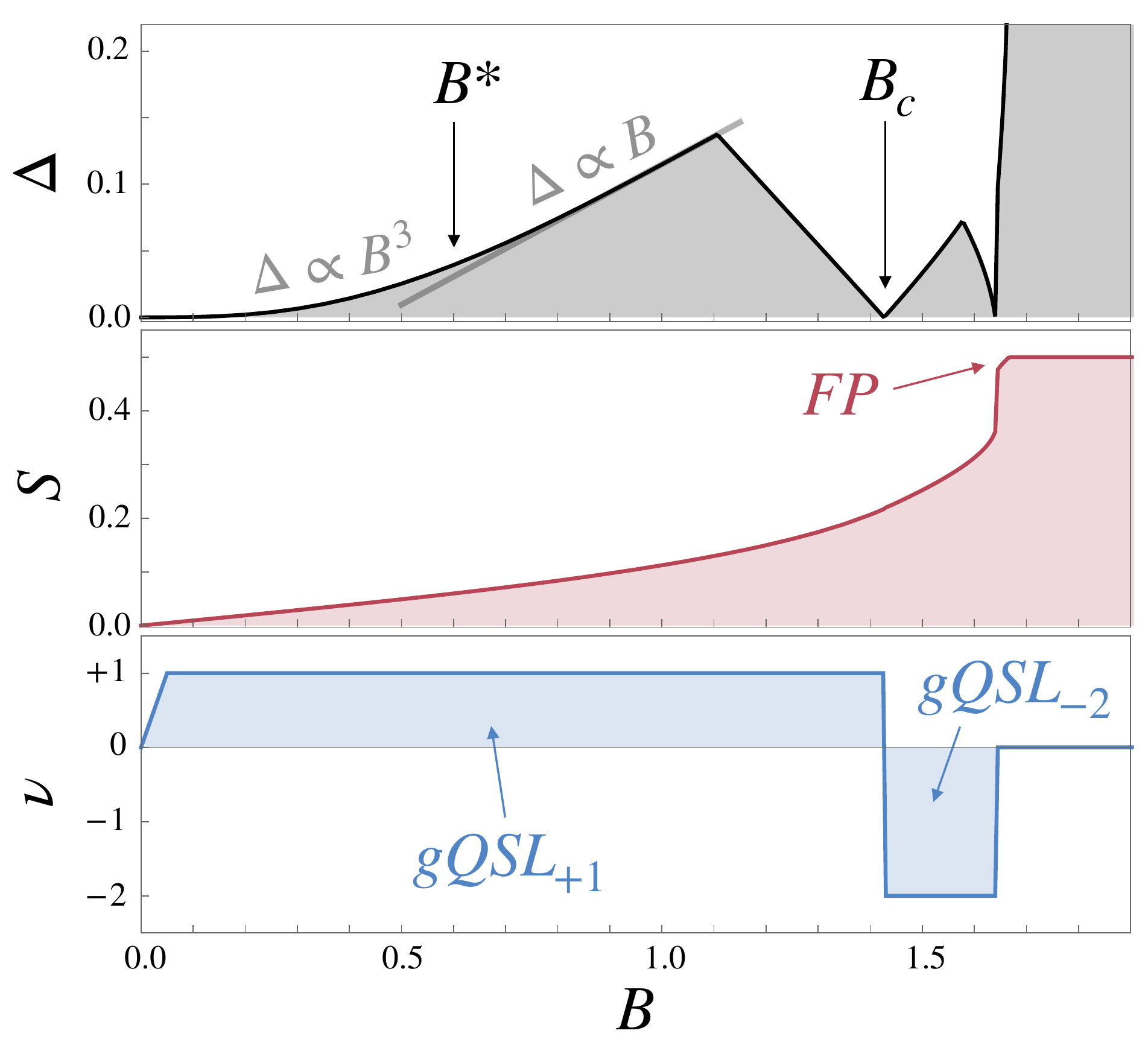} 
\caption{Dependence of the gap (top), the magnetic moment (middle) and the
		Chern number (bottom) with an applied magnetic field and zero DM interaction. The gap obtained
		from the MMFT changes from $\Delta \propto B^3$, expected from
		perturbation theory, to $\Delta \propto B$ around $B^* \sim 0.6$ under
		the magnetic field ${\bf B}=B(1,1,1)/\sqrt{3}$.  A topological
		transition from a gapped QSL with Chern number, $\nu=+1$, gQSL$_{+1}$,
		to a QSL with $\nu=-2$, gQSL$_{-2}$, occurs at $B_c \sim 1.43$. For $B
		\gtrsim  1.64$ a topologically trivial polarized insulator is
		stabilized. $K= 1$ in this plot.} \label{fig:fig4}
\end{figure}

To conclude, we discuss our results in the context of Kitaev materials.
Although they are mostly believed to have FM couplings
\cite{jackeli,winter,valenti}, some works suggest \cite{kee,banerjee2017} AFM
couplings as mostly considered here. QSL behavior has recently been observed in
the honeycomb magnet H$_3$LiIr$_2$O$_6$ \cite{kitagawa} with the caution
that H disorder in this material can deviate magnetic couplings from the 
Kitaev model \cite{li,vandenbrink}. 
Although $\alpha$-RuCl$_3$ is magnetically ordered, there is experimental evidence for
its proximity to a QSL phase \cite{banerjee2016,banerjee2017}. Under high
pressure above 1 GPa \cite{wang2019} or applying a magnetic field destroys AF
order giving way to a gapped QSL \cite{hentrich2018,baek2017}. Strikingly,
recent thermal Hall conductivity experiments find fractional quantization of
the thermal conductance which is attributed to the Majorana edge modes in a
gQSL$_{+1}$ \cite{motome}. NMR experiments find a spin gap $\Delta \propto B^3$
at small fields \cite{ruegg} due to the fractionalization of the spin into two
gauge fluxes and a gapped Majorana fermion as predicted by Kitaev
\cite{kitaev}.  Fig.~\ref{fig:fig4} shows that a similar spin gap opening with
a [111] magnetic field and zero DM should be observed in AFM Kitaev materials.
 Fig.~\ref{fig:fig4} also shows that in AFM Kitaev materials, the
gQSL$_{+1}$ (under a positive magnetic field) would survive up to an applied
tilted magnetic field of $B \lesssim B_c$, way beyond the perturbative regime.
The Majorana edge states in this QSL will contribute to the thermal Hall
conductance~\cite{kitaev,nagaosa2012}, $\kappa_{xy}/T = \frac{\pi}{12}
\frac{k_B^2}{\hbar d} $, as recently observed~\cite{motome}. This is half and
opposite in sign to the thermal Hall conductance observed in an integer quantum
Hall effect experiment \cite{kane1997} associated with electronic charge.
Strikingly, since a distinct gapped QSL with $\nu=-2$ in the range $B
\sim 1.43-1.64$ arises {($\sim$ 90 Tesla using $2 K \approx 7$ meV for $\alpha$-Ru Cl$_3$), our analysis predicts a sudden jump of $\kappa_{xy}/T$, the
thermal Hall coefficient, from $\frac{\pi}{12} \frac{k_B^2}{\hbar d} $ to
$-\frac{\pi}{6} \frac{k_B^2}{\hbar d} $ around $B_c$. 
This signals a novel topological transition in AFM Kitaev magnets that could be searched
experimentally.

{\it Acknowledgements.} J. M. acknowledges financial support from
(RTI2018-098452-B-I00) MINECO/FEDER, Uni\'on Europea, through 
the Mar\'ia de Maeztu Programme for Units of Excellence 
in R\&D (CEX2018-000805-M) and from (mobility program:
"Salvador de Madariaga": PRX18/00070) Ministerio de Educaci\'on, Cultura y
deporte in Spain and the hospitality from N\'eel Institute in Grenoble.
We thank anonymous referee for making us realize that the gap
closing to the fully polarized state was overlooked in the first version.

\section{Details on the Majorana mean field theory}

As mentioned in the main text, mapping the spins 1/2 onto four Majorana fermions can be seen as a two-step procedure. First we proceed to the standard parton construction by introducing 
two Abrikosov fermions:

\begin{eqnarray}
S_i^\alpha=\frac{1}{2} f_{i}^{a+} \sigma^\alpha_{ab} f_{i}^{b}.
\end{eqnarray}
These two fermion species carry the magnetic moment of the spin, and for each
of them, two Majorana fermions can be introduced as:

\begin{eqnarray}
f_{i\uparrow  }  &=&\frac{1}{2}(c_0 - i c_3) \nonumber \\
f_{i\uparrow  }^+&=&\frac{1}{2}(c_0 + i c_3) \nonumber \\
f_{i\downarrow}  &=&\frac{1}{2}(c_2 - i c_1) \nonumber \\
f_{i\downarrow}^+&=&\frac{1}{2}(c_2 + i c_1) 
\end{eqnarray}
with the specific commutation relation $\{c_i^a,c_j^b\} = 2 \delta_{ij}
\delta_{ab}$.  In the text, $c_0 = c$, $c_1 = b^x$, $c_2 = b^y$ and $c_3 =
b^z$, but let's keep the $c_i$ notation for simplicity. Defined like this, the spins read:

\begin{eqnarray}
S_i^x &=& \frac{1}{2}\left[ f_{i\uparrow}^+ f_{i\downarrow} + f_{i\downarrow}^+ f_{i\uparrow}\right]  = \frac{1}{4} ( +i c_1 c_0 + i c_3 c_2 ) \nonumber \\
S_i^y &=& \frac{-i}{2}\left[ f_{i\uparrow}^+ f_{i\downarrow} - f_{i\downarrow}^+ f_{i\uparrow}\right] = \frac{1}{4} ( +i c_2 c_0 + i c_1 c_3 ) \nonumber \\
S_i^z &=& \frac{1}{2}\left[ f_{i\uparrow}^+ f_{i\uparrow} - f_{i\downarrow}^+ f_{i\downarrow}\right]  = \frac{1}{4} ( +i c_3 c_0 + i c_2 c_1 )
\end{eqnarray}
and the single occupation constraints $f_{i\uparrow}^+ f_{i\uparrow} +
f_{i\downarrow}^+ f_{i\downarrow} = 1$, $f_{i\uparrow}^+ f_{i\downarrow}^+ = 0$
and $f_{i\uparrow} f_{i\downarrow}  =0$ (see below for discussion) yield to three additional terms to be introduced with corresponding Lagrange multipliers $\{ \lambda_i\}$ in the model:
\begin{eqnarray}
c_3c_0 - c_2c_1 = 0, \nonumber \\
c_2c_0 - c_1c_3 = 0, \nonumber \\
c_1c_0 - c_3c_2 = 0.
\label{constr}
\end{eqnarray}
Let's now focus on the pure isotropic Kitaev model for illustrating how the method is implemented. Injecting the above defined mapping and considering the constraints to be fulfilled, the Hamiltonian can be cast as:
\begin{eqnarray}
H_K &=& K \sum_{i \in u} \left[ S_i^x S_{i+d_1}^x + S_i^y S_{i+d_2}^y + S_i^z S_{i+d_3}^z \right] \\
      &=& -K \sum_{i \in u} \sum_{a=1}^{3}  \frac{i}{2} c^a_i c^a_{i+d_a} \frac{i}{2} c^0_i c^0_{i+d_a}
\end{eqnarray}
This Hamiltonian can now be mean-field decoupled following three different channels of binilears: 
\begin{eqnarray}
(a) &\to& \langle \frac{i}{2} c^a_i c^a_{i+d_a}  \rangle \frac{i}{2} c^0_i c^0_{i+d_a} + \frac{i}{2} c^a_i c^a_{i+d_a}  \langle \frac{i}{2} c^0_i c^0_{i+d_a}\rangle \nonumber \\
(b) &\to& - \langle \frac{i}{2} c^a_i  c^0_i \rangle \frac{i}{2} c^a_{i+d_a} c^0_{i+d_a}   - \frac{i}{2} c^a_i  c^0_i  \langle \frac{i}{2} c^a_{i+d_a} c^0_{i+d_a} \rangle \nonumber \\
(c) &\to& \langle \frac{i}{2} c^a_i  c^0_{i+d_a} \rangle  \frac{i}{2} c^a_{i+d_a} c^0_i + \frac{i}{2} c^a_i  c^0_{i+d_a} \langle \frac{i}{2} c^a_{i+d_a} c^0_i\rangle
\end{eqnarray}
and corresponding constant terms.
The mean field Hamiltonian is then Fourier transformed with the specific form:
\begin{eqnarray}
    c_i^a = \frac{\sqrt{2}}{\sqrt{n_s}} \sum_q e^{i q r_i} c_q^a
\end{eqnarray}
where $n_s$ is the number of Bravais sites in the lattice and the extra
$\sqrt{2}$ allows for recovering standard fermionic commutation relations in $q$-space:
\begin{eqnarray}
\sum_i \{ c_i^a c_i^a \} &=& \sum_{q} c_q^ac_{-q}^a + c_{-q}^ac_{q}^a = n_s \nonumber \\
\sum_q c_q^a c_q^a &=& 0, 	
\end{eqnarray}
where $c_{-q}^{a}$ can be identified to the creation operator $c_q^{a+}$. Thus, in the reciprocal space, the 
Majoranas follow the rules $\{ c_q^a , c_{q'}^b \}= \{ c_q^{a+} , c_{q'}^{b+} \} = 0$ and $\{ c_q^a , c_{q'}^{b+}\} =
\delta_{q,q'} \delta_{a,b}$ as standard fermions. 
Then, one can simply diagonalize the mean field Hamiltonian with all terms
decoupled as previously shown in the single particle basis. The set of Lagrange
multipliers $\{ \lambda_i \}$ are fixed by imposing the single occupancy
constraint given by Eq.~(\ref{constr}), at the mean-field level only, using
a least square minimization. This done, we obtain the mean field many-body wave
function by constructing a Slater determinant up to half-filling. The operators
of the type  $\frac{i}{2} c_i^a c_j^b$ are then computed in this ground state
and re-injected in the mean field hamiltonian leading to a set of self-consistent
equations (SCE) which are solved numerically without imposing any restrictions on the
mean-field solutions. We repeat this procedure until convergence of the mean field
parameters and the energy up to a desired tolerance $\delta$ is achieved. In
our case, we have fixed this tolerance to at least $\delta = 10^{-10}$ on the
parameters. Calculations are performed
following a Broyden minimization method \cite{broyden} up to
$60 \times60$ clusters which are sufficiently large to describe the behavior of
the model in the  thermodynamic limit.

All mean field parameters are in the ground state Slater determinant 
of the Majoranas, providing an efficient and stable way to converge towards the
saddle points. Our construction easily deals with enlarged unit-cells (we have 
checked our results up to 12 site unit-cells) which allows for higher symmetry breaking mean-
field solutions which could be overlooked in too small clusters. This is for
instance the case of the {\it zig-zag} and {\it stripy} phases reported in
the next section when the Heisenberg exchange term is added to the system.

In contrast to other Majorana mean-field theories, our approach 
imposes the single occupancy constraint on average throughout the full phase diagram. 
The differences between our approach and a close mean-field theory\cite{nasu} which uses
 a two-Majorana representation per site is provided in Section.~V of the present 
Supplementary Material together with the comparison with alternative approaches based on 
Abrikosov fermions \cite{schaffer,liu}. 
The possibility of explicitly having four occupied bands as in our MMFT, three associated with the gauge fluxes and one with the matter provides more
insights about the nature of the different phases in contrast to having only two occupied bands \cite{nasu}. An augmented parton construction has also been
proposed\cite{knolle}, in a model in which the constraint has not been violated. This approach 
is designed for the description of the dynamics of the excitations above the ground state energy 
in the Kitaev $+\Gamma$ term.

\section{Heisenberg-Kitaev Phase diagram}

As mentioned in the main text, the magnetic orders observed in candidate
materials of the Kitaev model indicate the presence of Heisenberg like
interactions and even other types of exchanges. Before studying in detail the
phase diagram under a magnetic field and the DM interaction, we have verified
that our self-consistent MMFT  with constraints can reproduce the phase diagram
of the Heisenberg-Kitaev model found through exact numerical
techniques\cite{chaloupcka,trebst2011}.
\begin{figure}[h] \includegraphics[width=0.25\textwidth,clip]{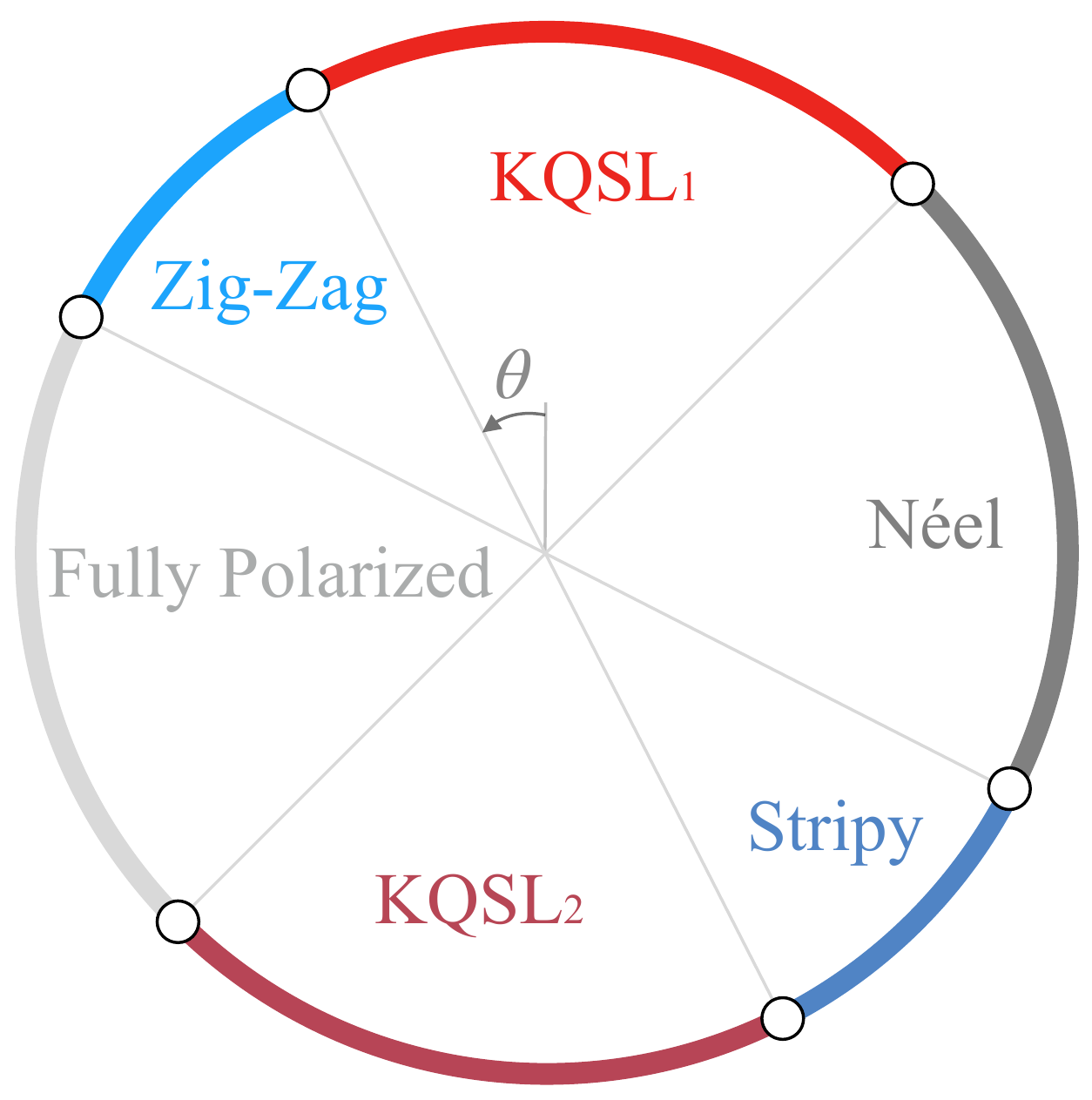}
\caption{Phase diagram of the Heisenberg-Kitaev model
obtained from the MMFT as a function of $\theta$. The extension of the two
Kitaev QSL, labeled KQSL$_1$ and KQSL$_2$ is very large. The pure Kitaev point
at $\theta=0$ is in the middle of KQSL$_1$. Different magnetic orders,
compatible with those already reported in the literature, are obtained by
enlarging the number of sites per unit cell in our theory, {\it e.g.} the
Zig-Zag and the Stripy phases.  } \label{fig:fig1supp} \end{figure}

Thus, we have considered the hamiltonian $H = H_K + H_J$ with $H_K = 2 K
\sum_{\langle i,j \rangle,\gamma} S_i^\gamma S_j^\gamma$ and $H_J = J
\sum_{\langle i,j \rangle} {\bf S}_i \cdot {\bf S}_j$. We then introduce the
extrapolation parameter $\theta$ defined as $\tan \theta = J/K$ and solve the
self-consistent equations of our theory for $\theta \in [0,2\pi]$. The
corresponding phase diagram is displayed in Fig.\ref{fig:fig1supp}. Note that
we have considered theories up to four sites per unit cell, necessary to
stabilize large unit-cell magnetic orders such as the Zig-Zag and the Stripy
phases \cite{lesphases}.

The good agreement with other techniques indicates the relevance of the method
and its capability of treating magnetic orders and QSL phases on equal footing.
In addition, we see that the pure Kitaev limit ($\theta=0$) lies deep in the
domain of the Kitaev QSL (1) (KQSL$_1$), and since we are interested in the
effects of the magnetic field and Dzyaloshinskii-Moriya interactions on this
QSL, we can restrict our analysis to this $\theta=0$ starting point without any
loss of generality.

\begin{figure}[h]
\includegraphics[width=0.35\textwidth,clip]{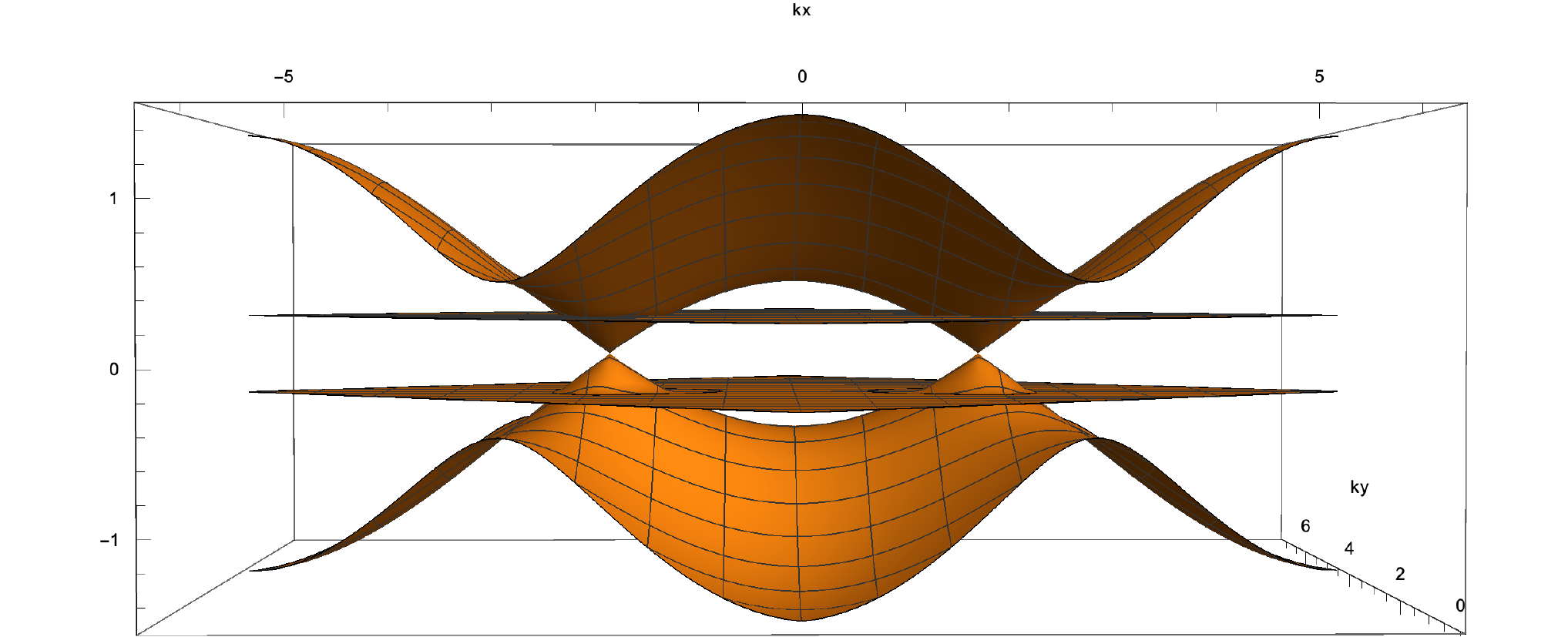}
\includegraphics[width=0.35\textwidth,clip]{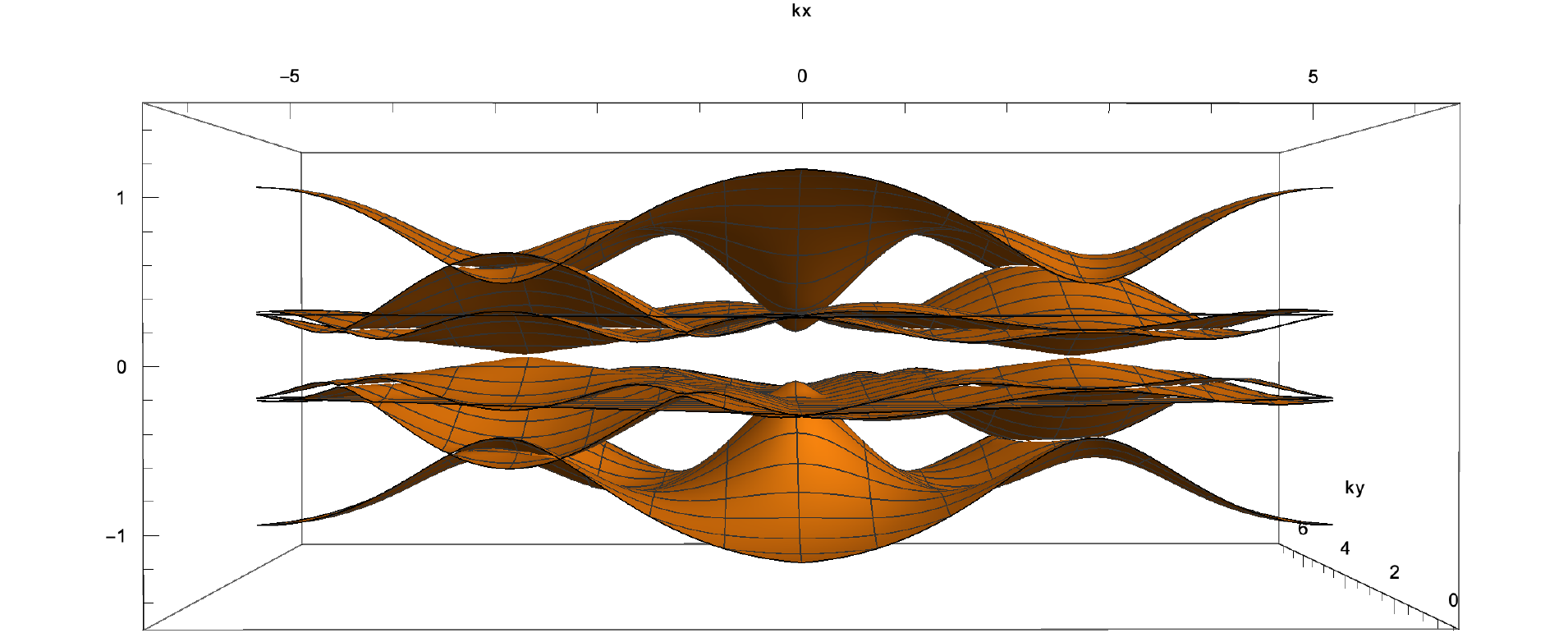}
\includegraphics[width=0.35\textwidth,clip]{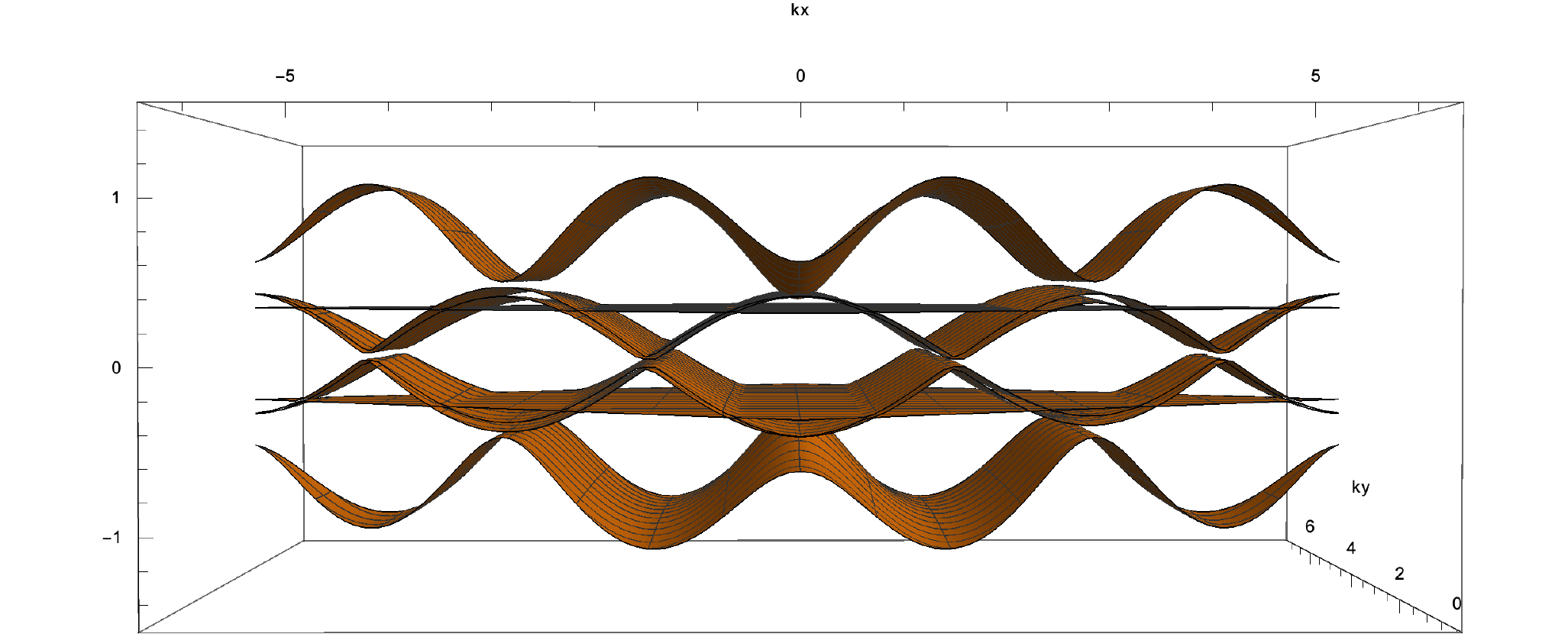} 
\caption{ Majorana dispersions of the Kitaev model in the
presence of the DM interaction. The dependence of the Majorana dispersions
with DM with no applied magnetic field, $B=0$, is shown. At a finite DM fo
$D=1.4$ (top panel) the Majorana dispersions resemble the bare dispersions of
the KQSL in which the linear Majorana dispersions and the flat flux bands are
evident. However, the Berry phases around the Dirac points are equal. As the DM
is increased up to $D=1.5$ a gap opens up in the spectrum and a gapped QSL with
$\nu=-1$, gQSL$_{-1}$ emerges. The whole spectrum of this state is strongly
modified due to strong hybridization between the flux and itinerant Majoranas
and the Dirac points located at $K$ and $K'$ in the pure Kitaev model ($D=0$)
move to new locations in the Brillouin zone.  The bottom panel shows the
dispersions at even larger values, $D=1.6$, in which the system is going into a
gapless state.  
  } 
\label{fig:figS2}
\end{figure}

\section{Effect of Dzyaloshinskii-Moriya interaction on Majorana spectrum}

As discussed in the main text, the DM interaction has an important effect on the ground state
of the pure Kitaev model {\it i. e.} the KQSL. Even with no applied magnetic field, a small but 
finite DM can induce a transition from a KQSL to an uQSL. The uQSL is a chiral QSL 
which is characterized by having the two Berry phases around the Dirac cones equal to $+ \pi$ or to $-\pi$
in contrast to the KQSL. 
At sufficiently large $D$ and with $B=0$ a transition to a gQSL$_{-1}$ (see the phase diagram shown in 
Fig.~1 of the main text) occurs. In order to analyze this transition we show in Fig.~\ref{fig:figS2} 
the change in the Majorana spectrum with increasing $D$.

\section{Multiband Berry phase and Chern number}
The Berry phases and Chern numbers determining the topological properties of
the model are obtained numerically using a multiband approach. This is
necessary since the itinerant and flux Majorana bands are occupied.  This is
done by defining the overlap matrix $M_{mn}^{({\bf k}_j,{\bf k}_{i})}=\langle
u_{m,{\bf k}_j} | u_{n,{\bf k}_i} \rangle$, with $m,n=1,...,N_b$ band indices
and ${\bf k}_i$ are discrete momenta of the first Brillouin zone. We define the
matrix $U({\bf k}_0)=\prod_{j=0}^{N-1} M_{mn}^{({\bf k}_j,{\bf k}_{j+1})}$
around a closed path in the BZ encircling a point ${\bf q}$ of interest, where
successive steps in momentum around the loop are taken. Note that ${\bf
k}_N={\bf k}_0$ for closing the loop. The Berry phase then reads $\phi({\bf
q})=\text{Im} \ln |U({\bf k}_0)|$. The Chern number is immediately obtained
from this expression by discretising the first Brillouin zone in elementary
four-site plaquettes \cite{fukui} (${\bf k}_j \to {\bf k}_j + {\bf u}_1  \to
{\bf k}_j + {\bf u}_1 + {\bf u}_2  \to {\bf k}_j + {\bf u}_2 \to {\bf k}_j $)
and defining the Berry flux matrix $F({\bf k}_j) =\text{Im} \ln \left| U({\bf
k}_j) \right|$ on each of them. The total Chern number $\nu$ of the system is
then nothing else but the sum of all the Berry fluxes over the whole BZ as
$\nu=\frac{1}{2\pi}\sum_j F({\bf k}_j)$ and so the thermal Hall coefficient is
half-quantized as: $\frac{\kappa_{xy}}{T} = \frac{\pi}{12} \frac{k_B^2}{\hbar
d} \nu$.

\section{Effect of the single-occupancy constraint}



One key feature of our Majorana parton approach stands in the appropriate implementation of the 
single particle constraint to deal with the Kitaev model under the magnetic field and the DM interaction. 
We now discuss the effect of the single occupancy constraint, needed to recover
the right spin operators from the Majorana fermions, on the magnetic properties of the
Kitaev model under an external magnetic field. Following \cite{kim}, in our MMFT we have imposed (on
average) the condition of single particle occupation leading to the following three constraints on the Majorana
fermions:
\begin{eqnarray}
\langle i b^x c \rangle - \langle i b^z b^y \rangle = 0,
\nonumber \\
\langle i b^y c \rangle-\langle i b^x b^z \rangle  = 0,
\nonumber \\
\langle i b^z c \rangle - \langle i b^y b^x \rangle =0.
\label{eq:constr}
\end{eqnarray}
This is in contrast with certain MMFT which do not impose these constraints
(on average) explicitly \cite{nasu} but assume that within the two Majorana representation
of the spin operators obtained through a Jordan Wigner transformation they are automatically 
satisfied. However, we show how this assumption is only valid in the pure Kitaev model. Within our MMFT we use a four Majorana
representation of the spins: $S^\alpha=i b^\alpha c$, which implicitly also assumes that the
single occupancy constraint given by the conditions over the Majorana fermions
$b^x c= b^z b^y, b^y c= b^x b^z, b^z c= b^y b^x$, is satisfied as in previous approaches. However, as
we discuss below, we have checked that the above constraints are only satisfied in the pure Kitaev model
{\it i.e.} when no magnetic field nor other exchange interactions are considered.
Hence, we found necessary in our MMFT to impose the constraints
(\ref{eq:constr}) in order to recover the one-particle condition per site 
at least on average.

\begin{figure}[h]
\includegraphics[width=0.25\textwidth,clip]{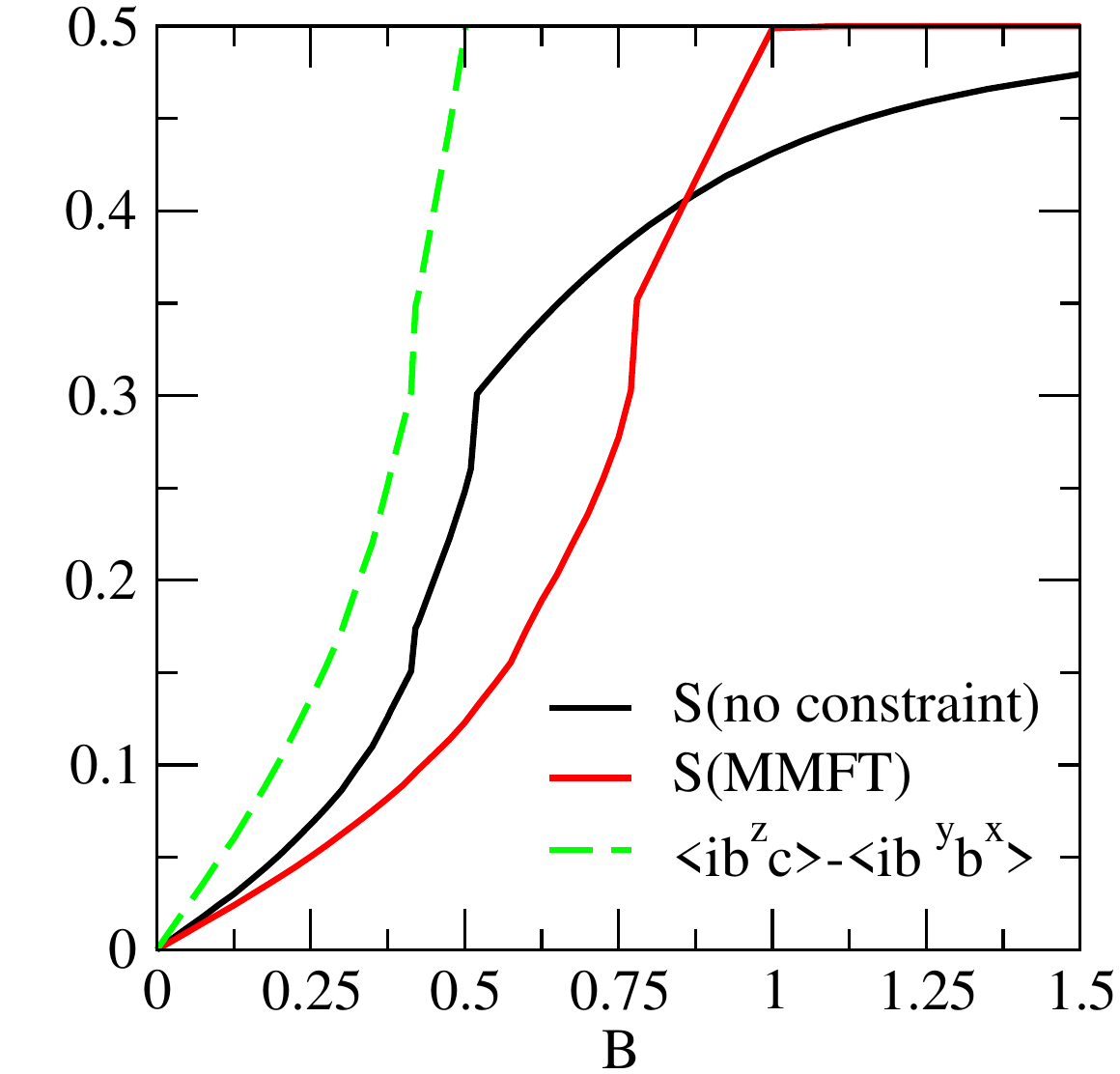}
\caption{Effect of removing the single occupancy constraint
on the dependence of the magnetic moment with an applied magnetic field. The
dependence of the magnetic moment, $S$, on a magnetic field in the $[001]$
direction applied to the Kitaev model with no DM interaction, $D=0$, is shown.
The results obtained from our MMFT with the constraints (\ref{eq:constr}) are
compared with the results in which no constraints are imposed. The dependence
of the constraint with the field shows how it is strongly violated as the
magnetic field is increased. $K=0.5$ in this plot. Calculations are performed
on a $36 \times 36$ cluster.} 
\label{fig:figS3}
\end{figure}

In Fig.~\ref{fig:figS3} we show results obtained within our MMFT with the
constraints (\ref{eq:constr}) fulfilled and compared with the case in which
they have been removed. The MMFT without constraints recovers the dependence of
the magnetic moment with a $[001]$ magnetic field reported previously in the
literature \cite{nasu}. Namely a transition from a gapless phase to another
gapless phase at which the $\pm \pi$ Berry phases associated with the two Dirac
cones switch their sign occurs around $B=0.42$ and a second transition to a
gapped polarized phase occurs at $B=0.52$.  When the constraints are imposed
though, these three different phases are present but their critical transition
values are shifted. Also we find that the polarized phase is fully saturated in
contrast to the case when no constraints are imposed. 

Let's discuss the effect of the constraints when a magnetic field is applied in
the $[111]$ direction. In Fig.~\ref{fig:figS4} we compare the magnetic moment
and the gap with and without the constraints imposed. The first important 
difference between the two approaches is that
while our MMFT (with the constraints imposed) naturally recovers the
topological gapped phase (with Chern number, $\nu=\pm 1$) even at small
magnetic fields $B$, as found by Kitaev. In contrast, the system is gapless up
to $B \sim 1.4$ in the no-constraint MMFT. In order to recover Kitaev's gapped
topological phase, three spin terms arising in perturbation theory to O($B^3$)
need to be added by hand to the Kitaev model \cite{nasu}. Hence, imposing the
constraint is readily essential for recovering the correct behavior of the
Kitaev model under a $[111]$ magnetic field in any 
regime of the magnetic field. Importantly, no topological phases are found
when the constraints are not imposed. While the two Berry phases around the
Dirac cones have opposite signs, $\pm \pi$, in the ungapped phase for $B
\lesssim 1.4$, the gapped phase found for $B \gtrsim 1.4$ is trivial, $\nu=0$,
as shown in Fig.~\ref{fig:figS4}.  Finally, without the constraints, there are n
signatures of the
$\nu=-2$ topological phase found with our MMFT  (Fig.~4 of the main text).
  
\begin{figure}[h]
\includegraphics[width=0.20\textwidth,clip]{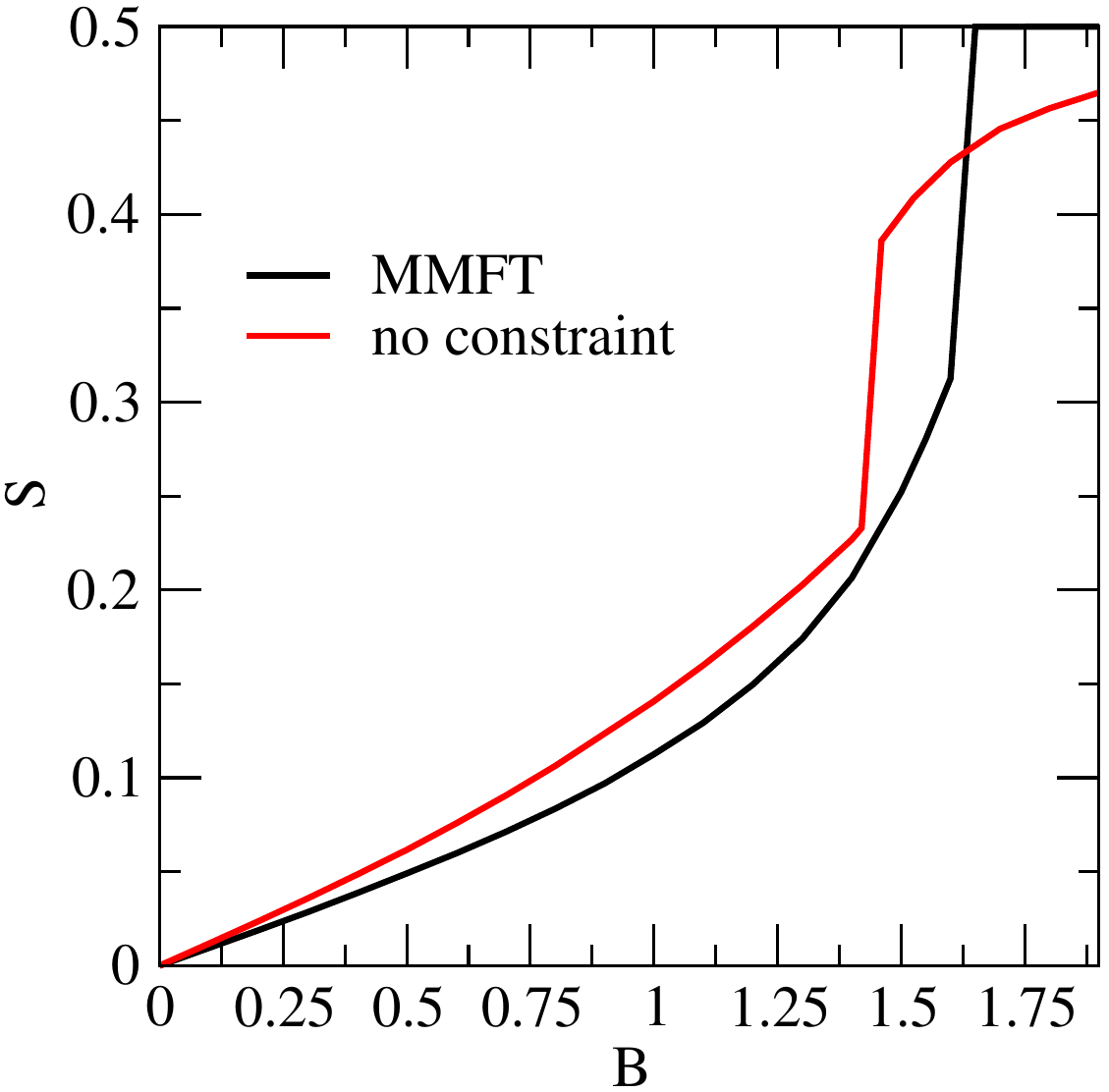}
\includegraphics[width=0.20\textwidth,clip]{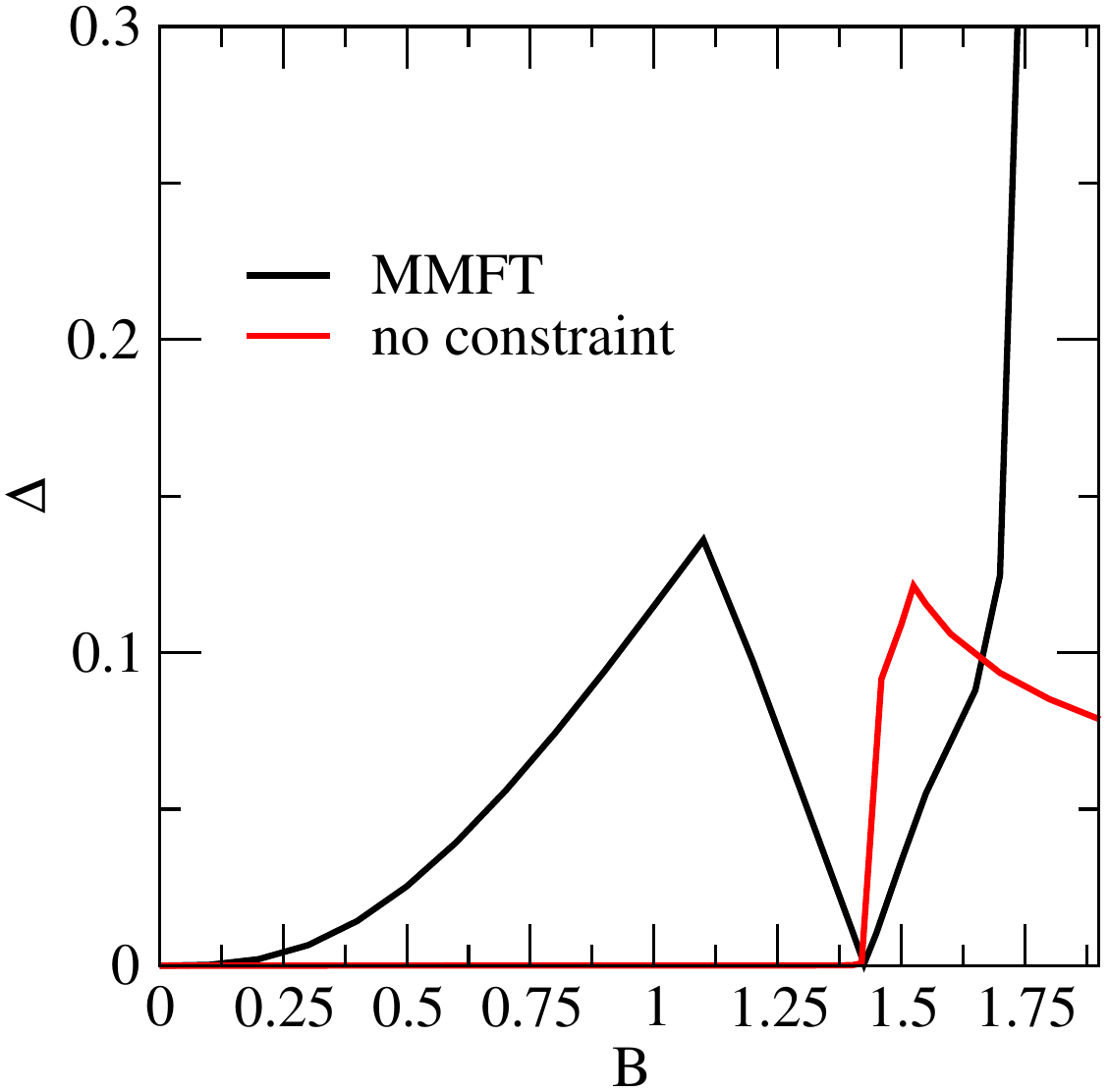}
\caption{ Effect of removing the single occupancy constraint
on the dependence of the magnetic moment and the gap with an applied magnetic
field. The dependence of the magnetic moment, $S$, and the gap on a $[111]$
magnetic field applied to the Kitaev model with no DM interaction, $D=0$, is
shown. The results obtained from the MMFT with the constraints
(\ref{eq:constr}) included are compared with the results in which no single
occupancy constraints imposed. No gap opening is found for $B \lesssim 1.4$
when the constraints are removed. $K=1$ in this plot. Calculations are
performed on a $36 \times 36$ cluster.} 
\label{fig:figS4}
\end{figure}

We finally discuss the relationship between our MMFT theory and mean-field 
decompositions based on Abrikosov fermions. \cite{gholke,liu} It has been shown \cite{kim} 
how a faithful Abrikosov fermion mean-field theory of the Kitaev model can be obtained after performing a 
unitary rotation on a Majorana mean-field hamiltonian similar to ours {\it i. e.} using the spin operators: $S_\alpha=b_\alpha c$, 
so the Majorana constraint $D=b_x b_y b_z c=1$ at each site is automatically satisfied 
with both spin liquid and magnetic channels taken into account. The derived Abrikosov fermion 
hamiltonian contains mean-field terms of the pairing type associated with the 
QSL channel and of the magnetic type as well as three 
constraints implying three Lagrange multipliers. Such mean-field construction not 
only leads to the correct exact KQSL and correct ground state energy when $B =0$ 
but also treats more accurately the QSL channels at non-zero $B$ 
as compared to conventional mean-field Abrikosov fermion
decompositions which introduce {\it only} one Lagrange multiplier (our $\lambda_z$) associated with the condition: 
$n_i=\sum_\sigma f^\dagger_{i\sigma} f_{i\sigma} =1$ (where $f^\dagger_i, f_i$ 
the Abrikosov fermions) on average~\cite{liu}.  Since, at the mean-field level, our MMFT and a conventional Abrikosov fermion approach
deal with the particle number constraint in different ways one should expect different results.  However, it 
is still interesting to note that a CSL with Chern number $\nu=\pm 2$ has been found in \cite{liu} when ${\bf B} || {\bf c}$ ({\it i. e.} along the $[111]$ direction) 
which is consistent with our gQSL$_{-2}$ phase although their model \cite{liu} contains an extra symmetric spin term 
that we do not include in our hamiltonian.

\section{Comparison to Exact Diagonalization}

\begin{figure}[ht]
\includegraphics[width=0.45\textwidth,clip]{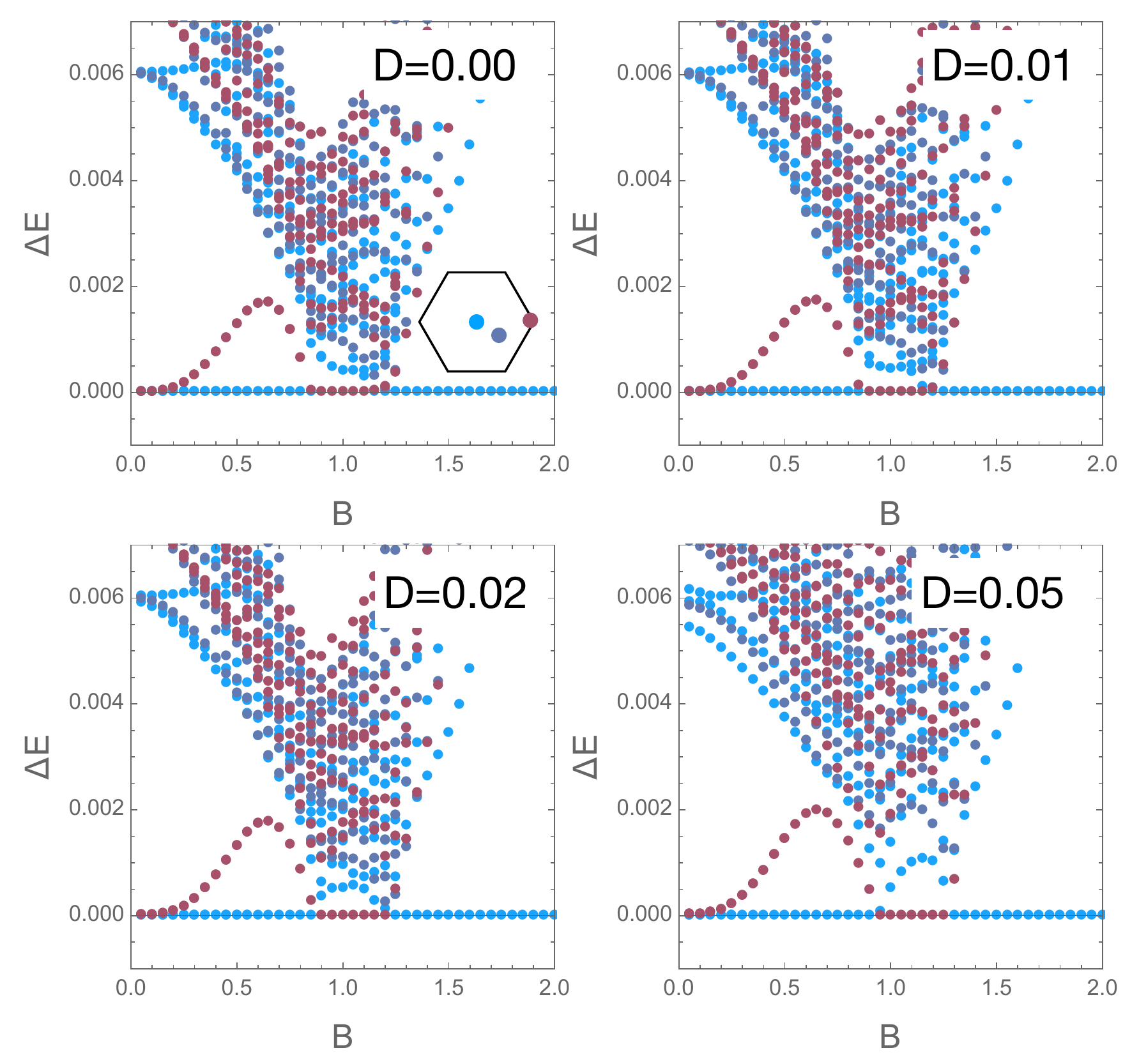}
\caption{ Gap spectra versus $B$ in the $[111]$ direction for various $D$
obtained by ED on a 18 site cluster. The first 10 levels are displayed in the
three inequivalent symmetry sectors (see inset) at the $\Gamma$ (light blue),
$X$ (dark blue) and $K$ (red) points. As $D$ increases from 0 to 0.05, the
gap in the intermediate phase, defined by the change of the ground state symmetry sector,
grows, in agreement with our MMFT results. Hence, gap opening in the thermodynamic 
limit cannot be excluded for any finite $D$.
} 
\label{fig:figS5}
\end{figure}

In order to challenge our gQSL$_{-2}$ phase, we have computed the gap spectra
of the model by exact diagonalizations (ED) on a 18 site cluster, for 4
different values of $D$, in function of $B$ along the $[111]$ direction and
using the translational symmetries (Fig.~\ref{fig:figS5}). As found in previous studies
\cite{trebst2019,trivedi}, the intermediate phase defined by the change of symmetry
in the ground state, is expected to be gapless, even though larger cluster sizes
should be analyzed to reach a definitive conclusion about the gap opening in the thermodynamic limit. 
However, for a cluster of 32 sites \cite{trebst2019}, a significant
shrinking of the energy excitations lets suggest that the gap might be
vanishing at the thermodynamic limit.

We are here interested on the behavior of the spectrum of the intermediate
phase under the presence of the DM term. As we have obtained in the MMFT, the
gQSL$_{-2}$ extends in the parameter space for finite K, D and B, and is always
gapped. In our ED results, starting from $D=0$ (upper-left panel) and slowly
increasing $D$, a clear increasing of the gap between the ground state energy
and the first excited state is observed. A finite size scaling would be
necessary for being definitely conclusive about a possible gap opening, but we
clearly show here that the combined effect of $D$ and $B$ tends to increase
the gap which is in good agreement with our MMFT.

The second effect of this combination is the location of the boundaries of the
intermediate phase. As $D$ increases, they are pushed to higher magnetic field,
a tendency that is also reproduced by our MMFT.

Since the gap we have found in the gQSL$_{-2}$ when $D=0$ is very small
in comparison with the one in the gQSL$_{+1}$ phase or the polarized phase,
it is not inconsistent with ED calculations, as these cannot exclude the presence of 
a tiny gap in the thermodynamic limit. Moreover, the behavior of the
intermediate phase (change of symmetry, $D$ vs. $B$, gap increasing) in ED
agrees with our phase diagram of Fig. 1 in our manuscript. This gives further support to the
existence of our intermediate gQSL$_{-2}$ phase, which should be gapped when 
$D$ and $B$ are combined \cite{lunkin}. As pointed out above, we
cannot exclude that, for $D=0$, the gap indeed closes in the thermodynamic limit
as suggested by exact numerical treatments on small systems \cite{trebst2019,trivedi}.

\section{Topological phase transitions from multiband Chern numbers}

\begin{figure}[ht]
\includegraphics[width=0.2\textwidth,clip]{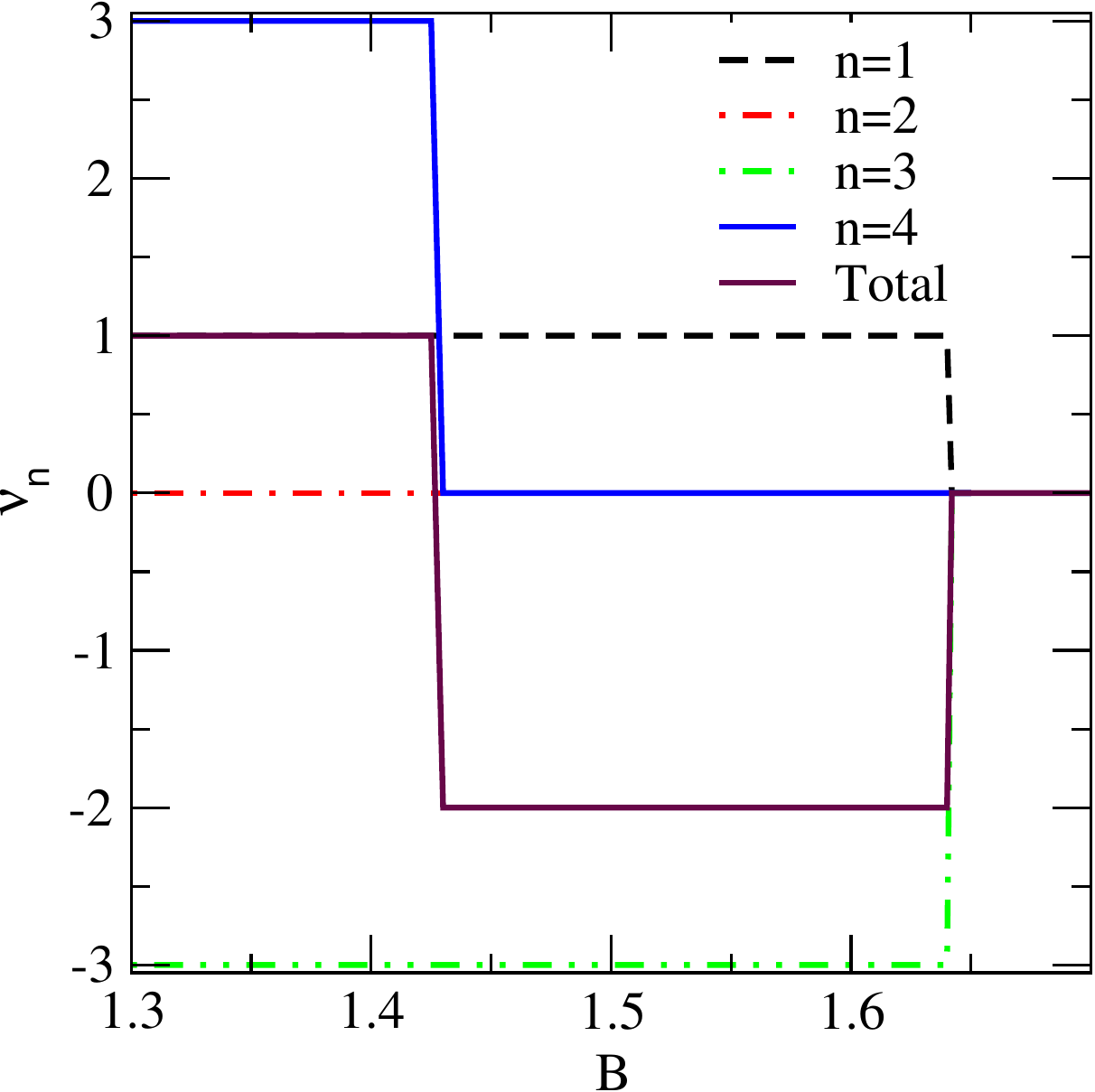}
\includegraphics[width=0.2\textwidth,clip]{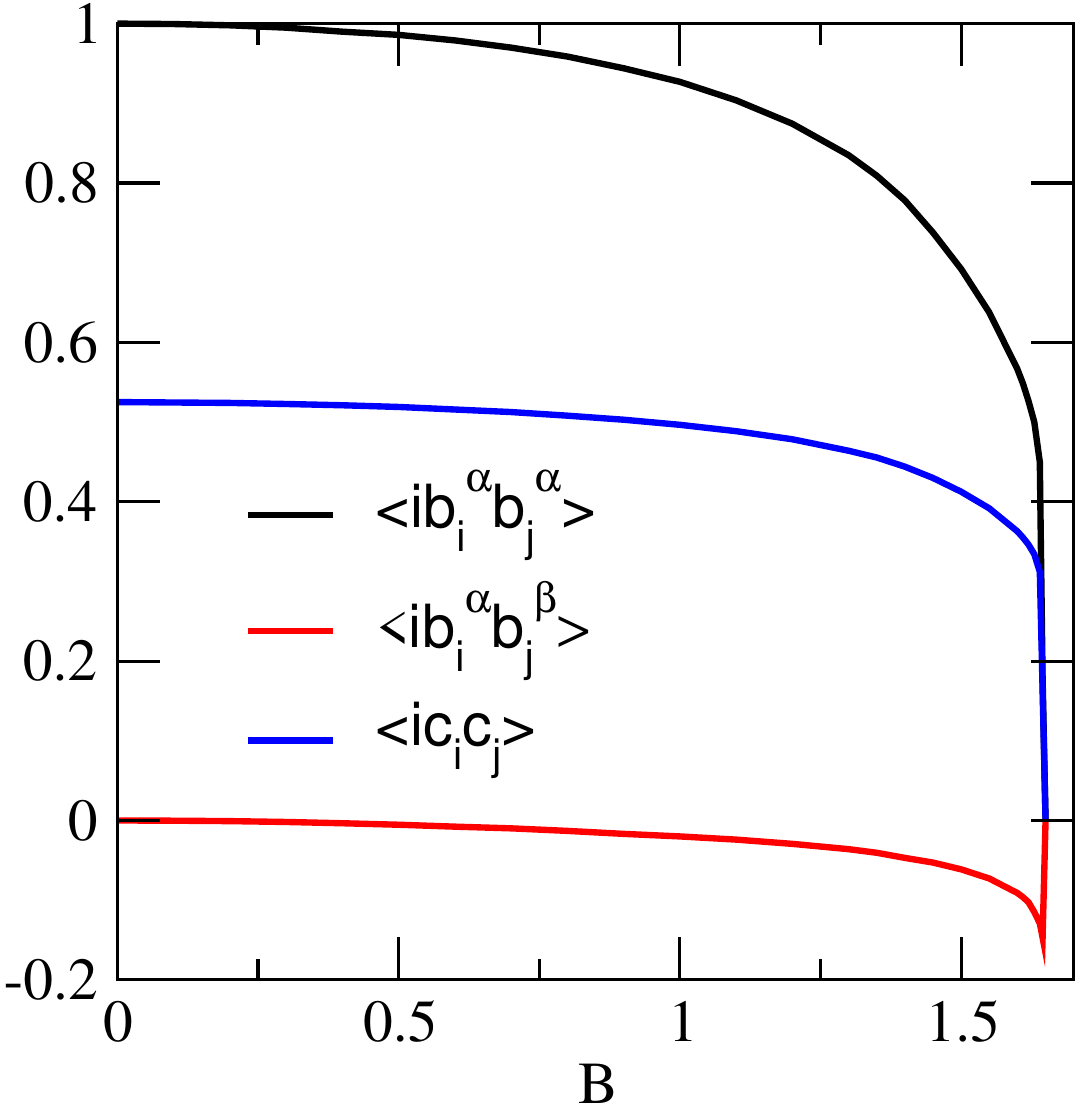}
\caption{Dependence of Chern numbers and Majorana bond averages on the magnetic field. In the
left plot the 
		decomposition of the total Chern number, $\nu$, into its band contributions $\nu_n$, for different applied magnetic fields, 
${\bf B}=B(1,1,1)/\sqrt{3}$, and no DM ($D=0$) are shown. The suppression of the
total Chern number, $\Delta \nu =-3$, is found to be solely due to the change in the Chern number 
in the fourth band, $\Delta \nu_4=-3$. In the right plot, the Majorana bond
averages as a function of $B$ are shown.} 
\label{fig:figS6}
\end{figure}

Further insight into the topological phase transitions occurring at $B_c=1.43$
and $B=1.64$ when no DM interaction is present (see Fig. 4 of
the main text) can be obtained by computing the Chern number, 
$\nu_n$, of the four lowest Majorana bands as shown in
Fig.~\ref{fig:figS6} for a system size of $72 \times 72$ sites. At
the phase transition from the gQSL$_{+1}$ to the gQSL$_{-2}$ at $B_c$ the drop
in the total Chern number, $\nu$ of 3 (with $\nu=\sum_n \nu_n$), can be
attributed solely to the change in the Chern number of the highest occupied
band from $\nu_4=3$ to $\nu_4=0$. Regarding the second transition from the
gQSL$_{-2}$ to the polarized phase, Fig.~\ref{fig:figS6} shows how at
$B=1.64$ the Chern numbers of all bands drop simultaneously to
zero, $\nu_n=0$, indicating the emergence of the polarized phase. In order to
rationalize the changes in the topological properties of the system we have
analyzed the dependence of the variational parameters with the applied field
${\bf B}=B(1,1,1)/\sqrt{3}$. The average Majorana bond values shown in
Fig.~\ref{fig:figS6}, $\langle i b^\alpha_i b^\alpha_j \rangle $, are
suppressed with $B$. For $B=0$,  $\langle i b^\alpha_i b^\beta_j \rangle = \delta_{\alpha,\beta}$,
recovering the exact zero-flux state of the localized Majorana fermions found
by Kitaev. As $B$ increases $ \langle i b^\alpha_i b^\alpha_j \rangle$ is only
slightly suppressed, $ \langle i b^\alpha_i b^\alpha_j \rangle \lesssim 1$ for
$ B \lesssim 0.5$. However, for larger $B > 1$ it is rapidly suppressed and
$\langle i b^\alpha_i b^\beta_j \rangle <0$ indicating that hybridization
between different Majoranas localized at the n.n sites occurs. The strong
suppression of $ \langle i b^\alpha_i b^\alpha_j \rangle$ with $B$ indicates
significant deviations from Kitaev's zero-flux state with the generation of 
non-zero $Z_2$ fluxes which become itinerant (see how the flat bands in Fig. 3 
of the main text gain dispersion with $B$). Hence, 
the jump of the Chern number around $B_c=1.43$ can be 
associated with such strong suppression. Indeed, the three localized bands,
$n=1,2,3$, associated with the bond Majorana fermions give zero contribution to
the Chern number: $\nu_1+\nu_2+\nu_3=0$ when $B=0$, as it should, since there
are no non-zero Z$_2$ fluxes present in the KQSL. However, the non-zero Z$_2$
fluxes generated as $B$ is increased, lead to an uncompensated contribution of
these bands to the total Chern number, $\nu_1+\nu_2+\nu_3 = -3$, and a total
Chern number, $\nu=-2$ when $B>B_c=1.43$. Although the matter Majorana bond
averages $\langle i c_ i c_j \rangle$ are also influenced by $B$ and
display a suppression with $B$ as shown in Fig. \ref{fig:figS6}, their
relevance on the topological transition and Chern number changes seems to be
only minor since these do not participate directly on the Z$_2$ flux dynamics.



\begin{thebibliography}{30}
\bibitem{balents} L. Balents, Spin liquids in frustrated magnets, Nature  {\bf 464}, 199 (2010).
\bibitem{savary} L. Savary and L. Balents, Quantum spin liquids: a review, Rep. Prog. Phys. {\bf 80}, 016502 (2017).
\bibitem{anderson1973} P. W. Anderson, Resonating valence bonds: A new kind of insulator?, Mater. Res. Bull. {\bf 8} (2): 153 (1973).
\bibitem{jackeli} G. Jackeli and G. Khalliulin, Mott Insulators in the Strong Spin-Orbit Coupling Limit: From Heisenberg to a Quantum Compass and Kitaev Models,  Phys. Rev. Lett. {\bf 102}, 017205 (2009).
\bibitem{takagi} H. Takagi, T. Takayama, G. Jackeli, and G. Khaliullin, Kitaev quantum spin liquid-concept and materialization, Nat. Rev. Phys. {\bf 1}, 264 (2019).
\bibitem{kitagawa} K. Kitagawa, {\it et. al.} A spin-orbital entangled quantum liquid on honeycomb lattice, Nature {\bf 554}, 341 (2018).
\bibitem{oshikawa2017} G. Y., Masashiko, H. Fujita, and M. Oshikawa Designing Kitaev spin liquids in metal-organic frameworks, Phys. Rev. Lett. 119, 057202 (2017). 
\bibitem{kitaev}  A. Kitaev, Anyons in an exactly solvable model and beyond,  Ann. Phys.  {\bf 321}, 2-111 (2006).
\bibitem{motome} Y. Kasahara, {\it et. al.}, Majorana quantization and half-integer thermal quantum Hall effect in a Kitaev spin liquid, Nature {\bf 559}, 227 (2018).
\bibitem{ruegg} N. Jansa, {\it et. al.} Observation of two types of fractional excitation in the Kitaev honeycomb magnet, Nat. Phys. {\bf 14}, 786 (2018).
\bibitem{yfjiang} Y. F. Jiang, T. P. Devereaux, and H.-C. Jiang, Field-induced quantum spin liquid in the Kiatev-Heisenberg model and its relation to $\alpha$-RuCl$_3$, Phys. Rev. B {\bf 100}, 165123 (2019).
 \bibitem{hcjiang} H.-C. Jiang,C.-Y. Wang, B. Huang, and Y.-M. Lu, Field induced quantum spin liquid with spinon Fermi surfaces in the Kitaev model, arXiv:1809.08247v2. 
 \bibitem{lzou} L. Zou, and Y.-C. He, Field-induced neutral Fermi surface and QCD$_3$-Chern-Simons quantum criticalities in Kitaev materials, arXiv:1809.09091v2.
 \bibitem{trebst} C. Kickey, and S. Trebst, Emergence of a field-driven U(1) spin liquid in the Kitaev honeycomb model, Nat. Comm. {\bf 10}, 1038 (2019). 
 \bibitem{kimchi} Z. Zhu, I. Kimchi,  D. N. Sheng, and L. Fu, Robust non-Abelian spin liquid and a possible intermediate phase in the antiferromagnetic Kitaev model with magnetic field, Phys. Rev. B {\bf 97}, 241110 (R) (2018).
 \bibitem{schaffer} R. Schaffer, S. Bhattacharjee, and Y. B.  Kim, Quantum phase transition in Heisenberg-Kitaev model, Phys. Rev. B {\bf 86}, 224417 (2012). 
 \bibitem{knolle}  J. Knolle, S. Bhattacharjee, and R. Moessner, Dynamics of a quantum spin liquid beyond integrability: The Kiatev-Heisenberg-$\Gamma$ model in an
 augmented parton mean-field theory, Phys. Rev. B {\bf 97}, 134432 (2018).
 \bibitem{gholke} M. Gohlke, G. Wachtel, Y. Yamaji, F. Pollmann, and Y. B. Kim, Quantum spin liquid signatures in Kitaev-like frustrated magnets, Phys. Rev. B {\bf 97}, 075126 (2018).
 \bibitem{valenti} S. M. Winter, Y. Li, H. O.  Jeschke, and R. Valent\'i, Challenges in design of Kitaev materials: Magnetic interactions from competing energy scales, Phys. Rev. B {\bf 93}, 214431 (2016).
\bibitem{lunkin} A. V. Lunkin, K. S. Tikhonov, M. V. Feigel'man, Perturbed Kitaev model: excitation spectrum and long-ranged spin correlations, Jour. of Phys. and Chem. of Sol. 128, 130, (2017).
\bibitem{RemarkOnHeisenbergTerm} We have mapped out the phase diagram of the
Heisenberg-Kitaev model confirming the presence of an extended KQSL phase (see
Supplementary material \cite{suppl}). Without loss of generality, we can thus restrict our
study to the pure Kitaev case.
\bibitem{perkins} N. B. Perkins, Y. Sizyuk, and P. W\"olfle.,  Interplay of many-body and single-particle interactions in iridates and rhodates, Phys. Rev. B {\bf 89}, 035143 (2014).
\bibitem{suppl} See Supplemental Material at http://link.aps.org/supplemental/ for (i) details of the Majorana Mean Field Theory, (ii) the extension to the Heisenberg-Kitaev model  (iii) the effect of the
DM interaction on the Majorana spectrum,  (iv) the computation of the multiband Berry phases and Chern numbers, and (v) the effect of the single-occupancy 
				constraint on magnetic properties of the model. It includes references \cite{chaloupka,fukui,broyden,trebst2011,lesphases}.
\bibitem{nasu} J. Nasu, Y. Kato,  Y. Kamiya, and Y. Motome, Succesive Majorana topological transitions driven by a 
magnetic field in the Kitaev model, Phys. Rev. B {\bf  98}, 060416 (R) (2018).
\bibitem{liu} Z.-X. Liu and B. Normand, Dirac and chiral quantum spin liquids on the honeycomb lattice in a magnetic field, Phys. Rev. Lett. {\bf 120}, 187201 (2018). 
\bibitem{choi} W. Choi, P. W. Klein, A. Rosch, and Y. B. Kim, Topological superconductivity in the Kondo-Kitaev model, Phys. Rev. B {\bf 98}, 155123 (2018).
\bibitem{yao} H. Yao, and  S. A. Kivelson, Exact chiral spin liquid with non-abelian anyons, Phys. Rev. Lett. {\bf 99}, 247203 (2007).
 \bibitem{trivedi} S. Pradhan, N. D. Patel, and N. Trivedi, Two-Magnon Bound States in the Kitaev model in a [111]-field, arXiv:1908.10877.
 \bibitem{winter} S. M. Winter, A. A. Tsirlin, M. Daghofer, J. van den Brink, Y. Singh, P. Gegenwart, and R. Valent\'i, Models and materials for generalized Kitaev magnetism, J. Phys.: Condens. Matter 29, 493002 (2017).
\bibitem{kee} H.-S. Kim, V. Shankar, A. Catuneanu, and H.-Y. Kee,  Kitaev magnetism in honeycomb RuCl3 with intermediate spin-orbit coupling, Phys. Rev. B {\bf 91}, 241110 (R) (2015).
\bibitem{banerjee2017} A. Banerjee, {\it et al.}, Neutron scattering in the proximate quantum spin liquid $\alpha$-RuCl$_3$, Science {\bf 356}, 1055 (2017).
\bibitem{li} Y. Li, S. M. Winter, and R. Valent\'i, Role of hydrogen in the spin-orbital-entangled quantum liquid candidate H$_3$LiIr$_2$O$_6$, Phys. Rev. Lett. {\bf 121}, 247202 (2018).
\bibitem{vandenbrink} R. Yadav, {\it et. al.}, Strong effect of hydrogen order on magnetic kitaev interactions in H$_3$LiIr$_2$O$_6$, Phys. Rev. Lett. {\bf 121}, 197203 (2018).
\bibitem{banerjee2016} A. Banerjee, {\it et. al.}, Proximate Kitaev quantum spin liquid behaviour in a honeycomb magnet, Nat. Mat. {\bf 15}, 733 (2016).
\bibitem{wang2019} Z. Wang,{\it et. al.}, Pressure-induced melting of magnetic order and emergence of a new quantum state in $\alpha$-RuCl$_3$, Phys. Rev. B {\bf 97}, 245149 (2019).
\bibitem{hentrich2018} R. Henritch, {\it et. al.}, Unusual phonon heat transport in $\alpha$-RuCl$_3$: strong spin-phonon scattering and field-induced spin gap, 
Phys. Rev. Lett. {\bf 120}, 117204 (2018).
\bibitem{baek2017} S.-H. Baek, S.-H. Do, K.-Y. Choi, Y. S. Kwon, A. U. B. Wolter, S. Nishimoto, J. van den Brink, and  B. B\"uchner, Evidence for a field-induced quantum spin liquid in $\alpha$-RuCl$_3$, Phys. Rev. Lett. {\bf 119}, 037201 (2017).
\bibitem{nagaosa2012} K. Nomura, S. Ryu,  A. Furusaki, and N. Nagaosa, Cross-correlated responses of topological superconductors and superfluids, Phys. Rev. Lett. {\bf 108}, 026802 (2012).
\bibitem{kane1997} C. L. Kane and M. P. A. Fisher, Quantized thermal transport in the fractional quantum Hall effect, Phys. Rev. B {\bf  55}, 15832 (1997).
%
\bibitem{broyden} J. Dennis, J. Mor\'e, Quasi-Newton Methods, Motivation and Theory, Society for Industrial and Applied Mathematics {\bf 19}, 46 (1977).
\bibitem{chaloupka} Chaloupka, J., Jackeli, G., and Khaliullin, G., Kitaev-Heisenberg Model on a Honeycomb Lattice: Possible Exotic Phases in Iridium Oxides A$_2$IrO$_3$, Phys. Rev. Lett. {\bf 105} 027204 (2010).
	\bibitem{trebst2011} H.-C. Jiang,  Z.-C. Gu, X.-L. Qi, and S. Trebst, Possible proximity of the Mott insulating iridate Na$_2$IrO$_3$ to a topological phase: Phase diagram of the Heisenberg-Kitaev model in a magnetic field, Phys. Rev. B {\bf 83}, 245104 (2011).
	\bibitem{lesphases}  D. Gotfryd, {\it et. al.}, Phase diagram and spin correlations of the Kitaev-Heisenberg model: Importance of quantum effects, Phys. Rev. B {\bf 95}, 024426 (2017).
\bibitem{fukui} Fukui, T., Hatsugai, Y., and Suzuki, H., Chern numbers and discretized Brillouin zone: efficient method of computing (spin) Hall conductances, 
Jour. of Phys. Soc. Jpn. {\bf 74}, 1674 (2005).

\end{thebibliography}

\begin{thebibliography}{30}
	\bibitem{broyden} J. Dennis, J. Mor\'e, Quasi-Newton Methods, Motivation and Theory, Society for Industrial and Applied Mathematics {\bf 19}, 46 (1977).
	\bibitem{nasu} J. Nasu, Y. Kato, Y. Kamiya, and Y. Motome, Successive Majorana topological transitions driven by a magnetic field in the Kitaev model, Phys. Rev. B {\bf 98}, 060416 (R) (2018).
	\bibitem{schaffer} R. Schaffer, S. Bhattacharjee, and Y. B.  Kim, Quantum phase transition in Heisenberg-Kitaev model, Phys. Rev. B {\bf 86}, 224417 (2012). 
	\bibitem{liu} Z.-X. Liu and B. Normand, Dirac and chiral quantum spin liquids on the honeycomb lattice in a magnetic field, Phys. Rev. Lett. {\bf 120}, 187201 (2018).
	\bibitem{knolle}  J. Knolle, S. Bhattacharjee, and R. Moessner, Dynamics of a quantum spin liquid beyond integrability: The Kitaev-Heisenberg-$\Gamma$ model in an augmented parton mean-field theory, Phys. Rev. B {\bf 97}, 134432 (2018).
	\bibitem{chaloupcka} J. Chaloupka, G. Jackeli, and G. Khaliullin, Kitaev-Heisenberg Model on a Honeycomb Lattice: Possible Exotic Phases in Iridium Oxides A$_2$IrO$_3$, Phys. Rev. Lett. {\bf 105} 027204 (2010).
	\bibitem{trebst2011} H.-C. Jiang,  Z.-C. Gu, X.-L. Qi, and S. Trebst, Possible proximity of the Mott insulating iridate Na$_2$IrO$_3$ to a topological phase: Phase diagram of the Heisenberg-Kitaev model in a magnetic field, Phys. Rev. B {\bf 83}, 245104 (2011).
	\bibitem{lesphases}  D. Gotfryd, {\it et. al.}, Phase diagram and spin correlations of the Kitaev-Heisenberg model: Importance of quantum effects, Phys. Rev. B {\bf 95}, 024426 (2017).
	\bibitem{fukui} T. Fukui, Y. Hatsugai, and H. Suzuki, Chern numbers and discretized Brillouin zone: efficient method of computing (spin) Hall conductances, Jour. of Phys. Soc. Jpn. {\bf 74}, 1674 (2005).
	\bibitem{kim} W. Choi, P. W. Klein, A. Rosch, and Y. B. Kim, Topological superconductivity in the Kondo-Kitaev model, Phys. Rev. B {\bf 98}, 155123 (2018).
	\bibitem{gholke}  M. Gholke, G. Wachtel, Y. Yamaji, F. Pollmann, and Y. B. Kim,  Quantum spin liquid signatures in Kitaev-like frustrated magnets, Phys. Rev. B {\bf 97}, 075126 (2018).
	\bibitem{trebst2019} C. Kickey, and S. Trebst, Emergence of a field-driven U(1) spin liquid in the Kitaev honeycomb model, Nat. Comm. {\bf 10}, 1038 (2019). 
	\bibitem{trivedi} S. Pradhan, N. D. Patel, and N. Trivedi, Two-Magnon Bound States in the Kitaev model in a [111]-field, arXiv:1908.10877.
	\bibitem{lunkin} A. V. Lunkin, K. S.  Tikhonov,  and M. V.  Feigel'man,  Perturbed Kitaev model: excitation spectrum and long-ranged spin correlations, Jour. of Phys. and Chem. of Sol. {\bf 128}, 130 (2017).
\end{thebibliography}
\end{document}